\documentstyle[12pt,bezier,twoside,makeidx]{report}
\newcommand{\dfrac}{\displaystyle \frac}
\newcommand{\msp}{\hspace{1.825ex}}
\newcommand{\Vsla}{V\hspace{-10.5pt} /}
\newcommand{\dsla}{\partial\hspace{-8pt} /}
\newcommand{\Dsla}{D\hspace{-9.25pt}  /}
\newcommand{\psla}{p\hspace{-6.25pt} /}
\newcommand{\ksla}{k\hspace{-7.5pt} /}
\makeindex
\begin{document}
\large
\bibliographystyle{plain}

\begin{titlepage}
\large
\hfill\begin{tabular}{l}HEPHY-PUB 621/95\\ UWThPh-1995-16\\ December 1995
\end{tabular}\\[3.8cm]
\begin{center}
{\Huge\bf EFFECTIVE POTENTIAL}\\[2ex]
{\Huge\bf MODELS FOR HADRONS}\\
\vspace{3.5cm}
{\Large\bf Wolfgang LUCHA}\\[.5cm]
Institut f\"ur Hochenergiephysik,\\
\"Osterreichische Akademie der Wissenschaften,\\
Nikolsdorfergasse 18, A-1050 Wien, Austria\\[1.5cm]
{\Large\bf Franz F.~SCH\"OBERL}\\[.5cm]
Institut f\"ur Theoretische Physik,\\
Universit\"at Wien,\\
Boltzmanngasse 5, A-1090 Wien, Austria\\[3.5cm]
\normalsize\it
Based on invited lectures (presented by F.~Sch\"oberl) at the\\
INTERNATIONAL SUMMER SCHOOL FOR STUDENTS ON\\
``DEVELOPMENT IN NUCLEAR THEORY AND PARTICLE PHYSICS''\\
Dubna, August 24 -- September 8, 1995\\
Supported by the Heisenberg--Landau Program\\
\end{center}
\end{titlepage}

\newpage

\thispagestyle{empty}
\vspace*{2cm}
\begin{center}
{\Large\bf Preface}
\end{center}

\noindent
The aim of these lectures is to give a self-contained introduction to
nonrelativistic potential models, to their formulation as well as to their
possible applications. At the price of some lack of (in a mathematical sense)
rigorous derivations, we try to give a feeling and understanding for the
simplest conceivable method to extract the explicit form of the forces acting
between quarks from the interplay between experimental observations and
theoretical considerations. According to this spirit, we demonstrate, in
detail, how to obtain the underlying Hamiltonian and how to determine the
Lorentz structure of the quark--(anti-)quark interaction potential from
well-established experimental facts.\\

\noindent
Vienna, December 1995\\

\hfill\begin{tabular}{l}{\sc Wolfgang Lucha}\\ {\sc Franz F.~Sch\"oberl}
\end{tabular}

\tableofcontents
\thispagestyle{empty}

\def\eexch{\setlength{\unitlength}{0.5mm}
\begin{picture}(60,160)
\thicklines
\put(0,0){\vector(0,1){40}}
\put(0,40){\line(0,1){39}}
\put(0,80){\circle{2}}
\put(0,81){\vector(0,1){44}}
\put(0,125){\line(0,1){35}}
\multiput(2,80)(10,0){6}{\line(1,0){6}}
\put(60,160){\vector(0,-1){40}}
\put(60,120){\line(0,-1){39}}
\put(60,80){\circle{2}}
\put(60,79){\vector(0,-1){44}}
\put(60,35){\line(0,-1){35}}
\put(-40,-10){\makebox(40,10)[tr]{$e^-(p_1,\sigma_1)\hspace{1ex}$}}
\put(60,-10){\makebox(40,10)[tl]{$\hspace{1ex} e^+(p_2,\sigma_2)$}}
\put(-40,160){\makebox(40,10)[br]{$e^-(q_1,\tau_1)\hspace{1ex}$}}
\put(60,160){\makebox(40,10)[bl]{$\hspace{1ex} e^+(q_2,\tau_2)$}}
\put(-40,75){\makebox(40,10)[r]{$e\,\gamma_\mu \hspace{1ex}$}}
\put(60,75){\makebox(40,10)[l]{$\hspace{1ex} e\,\gamma_\nu$}}
\put(10,65){\makebox(40,10)[t]{$\gamma$}}
\end{picture}}

\def\eann{\setlength{\unitlength}{0.5mm}
\begin{picture}(100,160)
\thicklines
\put(0,0){\vector(1,1){25}}
\put(25,25){\line(1,1){24}}
\put(50,50){\circle{2}}
\put(51,49){\vector(1,-1){28}}
\put(79,21){\line(1,-1){21}}
\multiput(50,52)(0,10){6}{\line(0,1){6}}
\put(100,160){\vector(-1,-1){25}}
\put(75,135){\line(-1,-1){24}}
\put(50,110){\circle{2}}
\put(49,111){\vector(-1,1){28}}
\put(21,139){\line(-1,1){21}}
\put(-40,-10){\makebox(40,10)[tr]{$e^-(p_1,\sigma_1)\hspace{1ex}$}}
\put(100,-10){\makebox(40,10)[tl]{$\hspace{1ex} e^+(p_2,\sigma_2)$}}
\put(-40,160){\makebox(40,10)[br]{$e^-(q_1,\tau_1)\hspace{1ex}$}}
\put(100,160){\makebox(40,10)[bl]{$\hspace{1ex} e^+(q_2,\tau_2)$}}
\put(50,50){\makebox(40,10)[bl]{$\hspace{1ex}e\,\gamma_\mu$}}
\put(50,100){\makebox(40,10)[tl]{$\hspace{1ex} e\,\gamma_\nu$}}
\put(50,75){\makebox(40,10)[l]{$\hspace{1ex}\gamma$}}
\end{picture}}

\def\eexchp{\setlength{\unitlength}{0.4719mm}
\begin{picture}(60,160)
\thicklines
\put(0,0){\vector(0,1){40}}
\put(0,40){\line(0,1){39}}
\put(0,80){\circle{2}}
\put(0,81){\vector(0,1){44}}
\put(0,125){\line(0,1){35}}
\multiput(2,80)(10,0){6}{\line(1,0){6}}
\put(60,160){\vector(0,-1){40}}
\put(60,120){\line(0,-1){39}}
\put(60,80){\circle{2}}
\put(60,79){\vector(0,-1){44}}
\put(60,35){\line(0,-1){35}}
\put(-40,-10){\makebox(40,10)[tr]{$e^-(p_1,\sigma_1)\hspace{1ex}$}}
\put(60,-10){\makebox(40,10)[tl]{$\hspace{1ex} e^+(p_2,\sigma_2)$}}
\put(-40,160){\makebox(40,10)[br]{$e^-(q_1,\tau_1)\hspace{1ex}$}}
\put(60,160){\makebox(40,10)[bl]{$\hspace{1ex} e^+(q_2,\tau_2)$}}
\put(-40,75){\makebox(40,10)[r]{$e\,\gamma_\mu \hspace{1ex}$}}
\put(60,75){\makebox(40,10)[l]{$\hspace{1ex} e\,\gamma_\nu$}}
\put(10,65){\makebox(40,10)[t]{$\gamma$}}
\end{picture}}

\def\eannp{\setlength{\unitlength}{0.4719mm}
\begin{picture}(100,160)
\thicklines
\put(0,0){\vector(1,1){25}}
\put(25,25){\line(1,1){24}}
\put(50,50){\circle{2}}
\put(51,49){\vector(1,-1){28}}
\put(79,21){\line(1,-1){21}}
\multiput(50,52)(0,10){6}{\line(0,1){6}}
\put(100,160){\vector(-1,-1){25}}
\put(75,135){\line(-1,-1){24}}
\put(50,110){\circle{2}}
\put(49,111){\vector(-1,1){28}}
\put(21,139){\line(-1,1){21}}
\put(-40,-10){\makebox(40,10)[tr]{$e^-(p_1,\sigma_1)\hspace{1ex}$}}
\put(100,-10){\makebox(40,10)[tl]{$\hspace{1ex} e^+(p_2,\sigma_2)$}}
\put(-40,160){\makebox(40,10)[br]{$e^-(q_1,\tau_1)\hspace{1ex}$}}
\put(100,160){\makebox(40,10)[bl]{$\hspace{1ex} e^+(q_2,\tau_2)$}}
\put(50,50){\makebox(40,10)[bl]{$\hspace{1ex}e\,\gamma_\mu$}}
\put(50,100){\makebox(40,10)[tl]{$\hspace{1ex} e\,\gamma_\nu$}}
\put(50,75){\makebox(40,10)[l]{$\hspace{1ex}\gamma$}}
\end{picture}}

\def\qexchp{\setlength{\unitlength}{0.4678mm}
\begin{picture}(60,160)
\thicklines
\put(0,0){\vector(0,1){40}}
\put(0,40){\line(0,1){39}}
\put(0,80){\circle{2}}
\put(0,81){\vector(0,1){44}}
\put(0,125){\line(0,1){35}}
\multiput(2,80)(10,0){6}{\line(1,0){6}}
\put(60,160){\vector(0,-1){40}}
\put(60,120){\line(0,-1){39}}
\put(60,80){\circle{2}}
\put(60,79){\vector(0,-1){44}}
\put(60,35){\line(0,-1){35}}
\put(-40,-10){\makebox(40,10)[tr]{$q_i(p_1,\sigma_1)\hspace{1ex}$}}
\put(60,-10){\makebox(40,10)[tl]{$\hspace{1ex} \bar q_j(p_2,\sigma_2)$}}
\put(-40,160){\makebox(40,10)[br]{$q_k(q_1,\tau_1)\hspace{1ex}$}}
\put(60,160){\makebox(40,10)[bl]{$\hspace{1ex} \bar q_\ell (q_2,\tau_2)$}}
\put(-40,75){\makebox(40,10)[r]{$g_{\rm s}\,\gamma_\mu\,
\dfrac{\lambda^a_{ki}}{2}\hspace{1ex}$}}
\put(60,75){\makebox(40,10)[l]{$\hspace{1ex} g_{\rm s}\,\gamma_\nu\,
\dfrac{\lambda^b_{j\ell}}{2}$}}
\put(10,65){\makebox(40,10)[t]{$G$}}
\end{picture}}

\def\qannp{\setlength{\unitlength}{0.4678mm}
\begin{picture}(100,160)
\thicklines
\put(0,0){\vector(1,1){25}}
\put(25,25){\line(1,1){24}}
\put(50,50){\circle{2}}
\put(51,49){\vector(1,-1){28}}
\put(79,21){\line(1,-1){21}}
\multiput(50,52)(0,10){6}{\line(0,1){6}}
\put(100,160){\vector(-1,-1){25}}
\put(75,135){\line(-1,-1){24}}
\put(50,110){\circle{2}}
\put(49,111){\vector(-1,1){28}}
\put(21,139){\line(-1,1){21}}
\put(-40,-10){\makebox(40,10)[tr]{$q_i(p_1,\sigma_1)\hspace{1ex}$}}
\put(100,-10){\makebox(40,10)[tl]{$\hspace{1ex} \bar q_j(p_2,\sigma_2)$}}
\put(-40,160){\makebox(40,10)[br]{$q_k(q_1,\tau_1)\hspace{1ex}$}}
\put(100,160){\makebox(40,10)[bl]{$\hspace{1ex} \bar q_\ell (q_2,\tau_2)$}}
\put(50,50){\makebox(40,10)[bl]{$\hspace{1ex}g_{\rm s}\,\gamma_\mu\,
\dfrac{\lambda^a_{ji}}{2}$}}
\put(50,108){\makebox(40,10)[tl]{$\hspace{1ex}g_{\rm s}\,\gamma_\nu\,
\dfrac{\lambda^b_{k\ell}}{2}$}}
\put(50,75){\makebox(40,10)[l]{$\hspace{1ex}G$}}
\end{picture}}

\def\trichter{\setlength{\unitlength}{0.48mm}
\begin{picture}(200,200)(0,-100)
\thinlines
\put(0,0){\circle{2}}
\put(1,0){\vector(1,0){189}}
\put(0,1){\vector(0,1){99}}
\put(0,-1){\line(0,-1){89}}
\bezier{301}(20,-80)(25.75,-31)(40,0)
\bezier{301}(40,0)(53,27)(80,40)
\bezier{401}(80,40)(120,60)(160,80)
\put(0,100){\makebox(0,0)[br]{$V(r)\hspace{1ex}$}}
\put(190,0){\makebox(0,0)[tl]{$\hspace{1ex}r$}}
\put(150,50){\vector(-1,2){6}}
\put(150,45){\makebox(0,0)[t]{$\hspace{1ex}V(r) \sim r$}}
\put(45,-65){\vector(-4,1){13}}
\put(45,-60){\makebox(0,0)[tl]{$\hspace{1ex}V(r) \sim \dfrac{1}{r}$}}
\end{picture}}

\def\vecprop{\setlength{\unitlength}{0.5mm}
\begin{picture}(0,0)(0,2)
\thicklines
\multiput(-18,0)(10,0){4}{\line(1,0){6}}
\put(-40,-5){\makebox(20,10)[r]{$\mu ,a\hspace{1ex}$}}
\put(20,-5){\makebox(20,10)[l]{$\hspace{1ex}\nu ,b$}}
\end{picture}}

\def\ferprop{\setlength{\unitlength}{0.5mm}
\begin{picture}(0,0)(0,10)
\thicklines
\put(-20,0){\vector(1,0){22}}
\put(2,0){\line(1,0){18}}
\put(-40,-5){\makebox(20,10)[r]{$i\hspace{1ex}$}}
\put(20,-5){\makebox(20,10)[l]{$\hspace{1ex}j$}}
\end{picture}}

\def\ghprop{\setlength{\unitlength}{0.5mm}
\begin{picture}(0,0)(0,10)
\thicklines
\multiput(-19,0)(11,0){3}{\line(1,0){5}}
\multiput(-11,0)(11,0){3}{\circle*{1}}
\put(14,0){\line(1,0){5}}
\put(-40,-5){\makebox(20,10)[r]{$a\hspace{1ex}$}}
\put(20,-5){\makebox(20,10)[l]{$\hspace{1ex}b$}}
\end{picture}}

\def\threevert{\setlength{\unitlength}{0.5mm}
\begin{picture}(0,0)(0,5)
\thicklines
\put(0,0){\circle{2}}
\multiput(1,-0.5)(10.95,-5.475){3}{\line(2,-1){7.3}}
\multiput(-1,-0.5)(-10.95,-5.475){3}{\line(-2,-1){7.3}}
\multiput(0,1)(0,10){3}{\line(0,1){6}}
\put(28,-14){\vector(2,-1){3}}
\put(-28,-14){\vector(-2,-1){3}}
\put(0,25){\vector(0,1){3}}
\put(-50,-21){\makebox(20,6)[tr]{$p,\mu ,a\hspace{1ex}$}}
\put(30,-21){\makebox(20,6)[tl]{$\hspace{1ex}q,\nu ,b$}}
\put(-10,30){\makebox(20,6)[b]{$r,\rho ,c$}}
\end{picture}}

\def\fourvert{\setlength{\unitlength}{0.5mm}
\begin{picture}(0,-2)
\thicklines
\put(0,0){\circle{2}}
\multiput(-2,2)(-11,11){3}{\line(-1,1){7.1}}
\multiput(2,2)(11,11){3}{\line(1,1){7.1}}
\multiput(2,-2)(11,-11){3}{\line(1,-1){7.1}}
\multiput(-2,-2)(-11,-11){3}{\line(-1,-1){7.1}}
\put(-50,24){\makebox(20,6)[tr]{$\mu ,a\hspace{1ex}$}}
\put(30,24){\makebox(20,6)[tl]{$\hspace{1ex}\nu ,b$}}
\put(30,-30){\makebox(20,6)[bl]{$\hspace{1ex}\rho ,c$}}
\put(-50,-30){\makebox(20,6)[br]{$\sigma ,d\hspace{1ex}$}}
\end{picture}}

\def\fervert{\setlength{\unitlength}{0.5mm}
\begin{picture}(0,0)(0,5)
\thicklines
\put(0,0){\circle{2}}
\put(30,-15){\vector(-2,1){15}}
\put(15,-7.5){\line(-2,1){14}}
\put(-1,-0.5){\vector(-2,-1){17}}
\put(-18,-9){\line(-2,-1){12}}
\multiput(0,1)(0,10){3}{\line(0,1){6}}
\put(-50,-21){\makebox(20,6)[tr]{$i\hspace{1ex}$}}
\put(30,-21){\makebox(20,6)[tl]{$\hspace{1ex}j$}}
\put(-10,30){\makebox(20,6)[b]{$\mu ,a$}}
\end{picture}}

\def\ghvert{\setlength{\unitlength}{0.5mm}
\begin{picture}(0,0)(0,5)
\thicklines
\put(0,0){\circle{2}}
\multiput(1,-0.5)(10.95,-5.475){3}{\line(2,-1){7.3}}
\multiput(-1,-0.5)(-10.95,-5.475){3}{\line(-2,-1){7.3}}
\multiput(10.125,-5.0625)(10.95,-5.475){2}{\circle*{1}}
\multiput(-10.125,-5.0625)(-10.95,-5.475){2}{\circle*{1}}
\multiput(0,1)(0,10){3}{\line(0,1){6}}
\put(-28,-14){\vector(-2,-1){3}}
\put(-50,-21){\makebox(20,6)[tr]{$p,a\hspace{1ex}$}}
\put(30,-21){\makebox(20,6)[tl]{$\hspace{1ex}b$}}
\put(-10,30){\makebox(20,6)[b]{$\mu ,c$}}
\end{picture}}

\def\combtad{\setlength{\unitlength}{0.5mm}
\begin{picture}(0,0)(0,-2.5)
\thicklines
\put(-25,-6){\line(0,1){12}}
\put(-23.15126,10.3425){\line(2,3){4.7}}
\put(-14.83034,20.53568){\line(2,1){7.3}}
\put(-4,25){\line(1,0){8}}
\put(7.53034,24.18568){\line(2,-1){7.3}}
\put(18.45126,17.3925){\line(2,-3){4.7}}
\put(25,6){\line(0,-1){12}}
\put(23.15126,-10.3425){\line(-2,-3){4.7}}
\put(14.83034,-20.53568){\line(-2,-1){7.3}}
\put(0,-25){\circle{2}}
\put(-7.53034,-24.18568){\line(-2,1){7.3}}
\put(-18.45126,-17.3925){\line(-2,3){4.7}}
\put(-1,-25){\line(-1,0){3}}
\put(1,-25){\line(1,0){3}}
\multiput(-8,-25)(-10,0){4}{\line(-1,0){6}}
\multiput(8,-25)(10,0){4}{\line(1,0){6}}
\end{picture}}

\def\combone{\setlength{\unitlength}{0.5mm}
\begin{picture}(0,0)(0,-2.5)
\thicklines
\put(-25,1){\line(0,1){5}}
\put(-25,0){\circle{2}}
\put(-25,-1){\line(0,-1){5}}
\put(-23.15126,10.3425){\line(2,3){4.7}}
\put(-14.83034,20.53568){\line(2,1){7.3}}
\put(-4,25){\line(1,0){8}}
\put(7.53034,24.18568){\line(2,-1){7.3}}
\put(18.45126,17.3925){\line(2,-3){4.7}}
\put(25,1){\line(0,1){5}}
\put(25,0){\circle{2}}
\put(25,-1){\line(0,-1){5}}
\put(23.15126,-10.3425){\line(-2,-3){4.7}}
\put(14.83034,-20.53568){\line(-2,-1){7.3}}
\put(4,-25){\line(-1,0){8}}
\put(-7.53034,-24.18568){\line(-2,1){7.3}}
\put(-18.45126,-17.3925){\line(-2,3){4.7}}
\multiput(-26,0)(-10,0){3}{\line(-1,0){6}}
\multiput(26,0)(10,0){3}{\line(1,0){6}}
\end{picture}}

\def\combtwo{\setlength{\unitlength}{0.5mm}
\begin{picture}(0,0)(0,-2.5)
\thicklines
\put(-25,1){\line(0,1){5}}
\put(-25,0){\circle{2}}
\put(-25,-1){\line(0,-1){5}}
\put(-23.15126,10.3425){\line(2,3){4.7}}
\put(-14.83034,20.53568){\line(2,1){7.3}}
\put(-4,25){\line(1,0){8}}
\put(7.53034,24.18568){\line(2,-1){7.3}}
\put(18.45126,17.3925){\line(2,-3){4.7}}
\put(25,1){\line(0,1){5}}
\put(25,0){\circle{2}}
\put(25,-1){\line(0,-1){5}}
\put(23.15126,-10.3425){\line(-2,-3){4.7}}
\put(14.83034,-20.53568){\line(-2,-1){7.3}}
\put(4,-25){\line(-1,0){8}}
\put(-7.53034,-24.18568){\line(-2,1){7.3}}
\put(-18.45126,-17.3925){\line(-2,3){4.7}}
\multiput(-26,0)(-10,0){3}{\line(-1,0){6}}
\multiput(26,0)(10,0){3}{\line(1,0){6}}
\multiput(-23,0)(10,0){5}{\line(1,0){6}}
\end{picture}}

\newpage

\setcounter{page}{1}

\chapter{Nonrelativistic
Potential}\label{ch:potential}\index{potential}\index{quark--quark potential}

In principle, the appropriate framework for the description of bound states
within relativistic quantum field theories is the Bethe--Salpeter formalism.
There are, however, some circumstances which are opposed to this. The
Bethe--Salpeter equation cannot be solved in general. The interaction kernel
entering in this equation is not derivable from QCD either. The propagators
of the constituents have to be approximated by their free form, the involved
masses, however, being interpreted as effective (``constituent'') ones. So,
even if one is willing to put up with the complexity of the Bethe--Salpeter
formalism, it is hard to obtain information from this approach.{\boldmath$ $}

The alternative which comes closest to one's physical intuition is the
description of bound states with the help of the Schr\"odinger equation
\cite{lucha89bi,lucha91,lucha92}
$$
H\,\psi = E\,\psi \ ,
$$
where the nonrelativistic Hamiltonian for a quantum system consisting of two
particles with masses $m_1$ and $m_2$, respectively, which interact via some
potential $V({\bf x})$ is given in the center-of-momentum frame by
$$
H = m_1 + m_2 + \dfrac{{\bf p}^2}{2\,\mu} + V({\bf x}) \ ;
$$
here $\mu$ denotes the reduced mass,
$$
\mu \equiv \frac{m_1\,m_2}{m_1 + m_2} \ .
$$

Our main task is simply to find that potential $V({\bf x})$ which describes
the interaction of the two particles constituting the bound state under
consideration. By investigating the corresponding scattering problem of the
involved bound-state constituents, the {\em perturbatively\/} accessible part
of this potential may be derived according to the following recipe (for
details see, for instance, Refs.~\cite{lucha91rel,lucha92com}):
\begin{enumerate}
\item Compute the scattering amplitude $T_{\rm f\/i}$, which is defined in
terms of the S-matrix element $S_{\rm f\/i}$ introduced in
Appendix~\ref{app:smatrix},
$$
S_{\rm f\/i} \equiv \langle{\rm f},\mbox{out}|{\rm i},\mbox{in}\rangle \ ,
$$
by the decomposition
$$
S_{\rm f\/i} = \delta_{\rm f\/i}
+ i\,(2\pi)^4\,\delta^{(4)}(P_{\rm f} - P_{\rm i})\,T_{\rm f\/i} \ ,
$$
for the {\em elastic\/} scattering process ${\rm i}\rightarrow{\rm f}$ in
lowest non-trivial order of perturbation theory, the so-called ``first Born
approximation.''
\item Perform the nonrelativistic limit, realized by the vanishing of the
momenta ${\bf p}$ of the involved bound-state constituents; we indicate this
limit rather symbolically by
$$
{\bf p} \rightarrow 0 \ .
$$
\item Obtain the configuration-space interaction potential sought after,
$V({\bf x})$, as the Fourier transform of the above scattering amplitude
$T_{\rm f\/i}$:
$$
V({\bf x})
= - (2\pi)^3 \int d^3k\exp(-i\,{\bf k}\cdot{\bf x})\,T_{\rm f\/i}(k) \ .
$$
\end{enumerate}

For the sake of simplicity, we split off all the normalization factors of the
one-particle wave functions, given for a fermion of mass $m$ and kinetic
energy $$E_p = \sqrt{{\bf p}^2 + m^2}$$ by
$$
\frac{1}{(2\pi)^{3/2}}\,\sqrt{\frac{m}{E_p}} \ ;
$$
we thereby define a quantity $t$ according to
$$
T_{\rm f\/i} =:
\frac{1}{(2\pi)^6}\,\frac{m^2}{\sqrt{E_{p_1}\,E_{p_2}\,E_{q_1}\,E_{q_2}}}\,t\ .
$$

In the framework of this nonrelativistic treatment, the short-range part of
the quark--antiquark potential (which is of perturbative origin!) will be
determined from quantum chromodynamics (QCD) according to the above
prescription. The shape of the long-range, confining part of the
quark--antiquark potential (which is of nonperturbative origin!) will be
obtained from the analysis of the possible Lorentz structures of the
potential, its coordinate dependence from the comparison of some of the
resulting predictions with experiment.

\section{Nonrelativistic limit}\index{nonrelativistic limit|(}

In the {\em nonrelativistic limit}, our whole formalism, so to say,
``collapses'' to an extremely simple one:
\begin{itemize}
\item The relativistically correct expression for the one-particle kinetic
energy,
$$
E_p = \sqrt{{\bf p}^2 + m^2} \ ,
$$
reduces to
$$
E_p = m \ .
$$
\item The Dirac spinors\index{Dirac spinor} $u(p,\sigma)$ and $v(p,\sigma)$
describing fermions of mass $m$, four-momentum $p$, and spin polarization
$\sigma$,
$$
\begin{array}{rcl}
u(p,\sigma) &=& \sqrt{\dfrac{E_p+m}{2\,m}}\left(\begin{array}{c}1\\[1ex]
\dfrac{\mbox{\boldmath$\sigma$}\cdot{\bf p}}{E_p+m}
\end{array}\right)\chi_\sigma\ ,\\
&&\\[-1ex]
v(p,\sigma) &=& \sqrt{\dfrac{E_p+m}{2\,m}}\left(\begin{array}{c}
\dfrac{\mbox{\boldmath$\sigma$}\cdot{\bf p}}{E_p+m}\\[2ex]
1\end{array}\right)\chi^c_\sigma\ ,\quad
\chi^c_\sigma \equiv -i\,\sigma_2\,\chi^\ast_\sigma\ ,
\end{array}
$$
where $\mbox{\boldmath$\sigma$} \equiv \{\sigma_i, i = 1,2,3\}$ are the three
Pauli matrices\index{Pauli matrices}
$$
\sigma_1 = \left(\begin{array}{cr}0&1\\ 1&0\end{array}\right)\ ,\quad
\sigma_2 = \left(\begin{array}{cr}0&-i\\ i&0\end{array}\right)\ ,\quad
\sigma_3 = \left(\begin{array}{cr}1&0\\ 0&-1\end{array}\right)\ ,
$$
reduce to the nonrelativistic spinors
\begin{eqnarray*}
u(\sigma) &=& \left(\begin{array}{c}\chi_\sigma\\ 0\end{array}\right)\ ,
\\[1ex]
v(\sigma) &=& \left(\begin{array}{c}0\\ \chi^c_\sigma\end{array}\right)\ .
\end{eqnarray*}
\item The normalization of the Dirac spinors\index{Dirac spinor!---,
normalization} adopted by us,
$$
u^\dagger(p,\sigma)\,u(p,\tau) = v^\dagger(p,\sigma)\,v(p,\tau)
= \frac{E_p}{m}\,\delta_{\sigma\tau}\ ,
$$
reduces to the nonrelativistic normalization
\begin{eqnarray*}
u^\dagger(\sigma)\,u(\tau) &=& \chi_\sigma^\dagger\,\chi_\tau
= \delta_{\sigma\tau}\ ,\\[1ex]
v^\dagger(\sigma)\,v(\tau) &=& {\chi^c_\sigma}^\dagger\,\chi^c_\tau
= \delta_{\tau\sigma}\ .
\end{eqnarray*}
\end{itemize}

\section{Static potential in quantum
electrodynamics}\label{sec:pot-qed}\index{quantum
electrodynamics|(}\index{QED|(}

Let us illustrate the very simple procedure outlined above by applying it
first to electron--positron scattering\index{electron--positron
scattering|(}:
$$
e^-(p_1,\sigma_1) + e^+(p_2,\sigma_2) \rightarrow
e^-(q_1,\tau_1) + e^+(q_2,\tau_2) \ .
$$
\begin{figure}[htb]
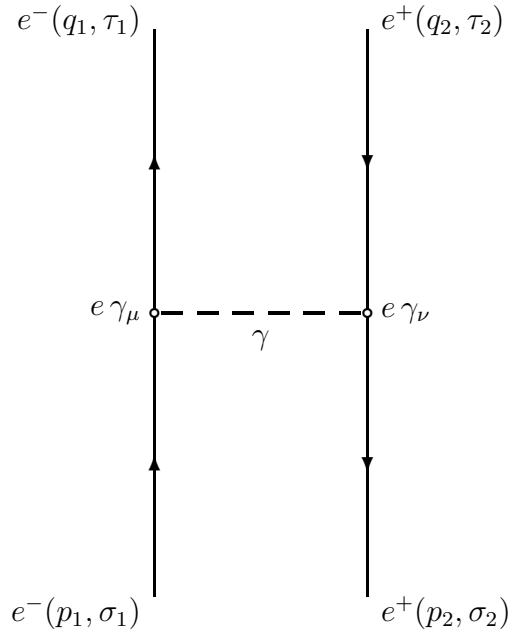

\normalsize
\begin{center}
\eexchp
\end{center}
\large
\caption{Electron--positron scattering, one-photon exchange
graph.}\index{one-photon exchange}\label{fig:elposscat-ex}
\end{figure}
\begin{figure}[htb]
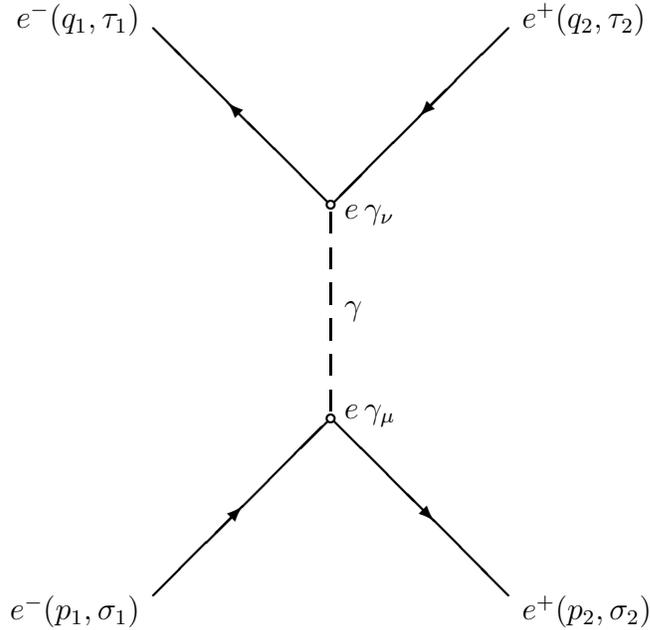

\normalsize
\begin{center}
\eannp
\end{center}
\large
\caption{Electron--positron scattering, pair annihilation
graph.}\index{pair annihilation}\label{fig:elposscat-ann}
\end{figure}

\noindent
The interaction term in the Lagrangian of quantum electrodynamics (QED) for
the coupling of a fermion with electric charge $e$, described by the Dirac
spinor field $\psi(x)$, to the photon field $A^\mu(x)$ is well known:
$$
{\cal L}_{\rm I}^{\rm QED}(x)
= e\,\bar\psi(x)\,\gamma_\mu\,\psi(x)\,A^\mu(x)\ .
$$

In lowest non-trivial order of the perturbative loop expansion, just two
Feynman diagrams contribute to the scattering amplitude $T_{\rm f\/i}$ for
elastic electron--positron scattering:
\begin{itemize}
\item the exchange\index{one-photon exchange} of a single photon, $\gamma$,
between electron and positron, as depicted in Fig.~\ref{fig:elposscat-ex};
and
\item the annihilation\index{pair annihilation} of the electron--positron
pair into a single photon, $\gamma$, followed by a subsequent creation of an
electron--positron pair by this single photon, as depicted in
Fig.~\ref{fig:elposscat-ann}.
\end{itemize}

With the help of the Feynman rules given for an arbitrary, that is, in
general, non-Abelian, gauge theory in Appendix~\ref{app:feynman}, it's
straightforward to find the corresponding scattering amplitude $T_{\rm f\/i}$:
\begin{itemize}
\item The contribution of the one-photon exchange\index{one-photon exchange}
graph in Fig.~\ref{fig:elposscat-ex} to our scattering amplitude $t$ reads
\begin{equation}
t_{\rm exch} = - \frac{e^2}{k^2}\,
\bar u(q_1,\tau_1)\,\gamma_\mu\,u(p_1,\sigma_1)\,
\bar v(p_2,\sigma_2)\,\gamma^\mu\,v(q_2,\tau_2)\ ,
\label{eq:t-rel}
\end{equation}
where $k$ denotes the involved momentum transfer,
$$
k \equiv p_1 - q_1 = q_2 - p_2\ .
$$
The square of this momentum transfer,
\begin{eqnarray*}
k^2 &\equiv& (p_1 - q_1)^2\\[1ex]
&=& (E_{p_1} - E_{q_1})^2 - {\bf k}^2\ ,
\end{eqnarray*}
which enters in the denominator of the scattering amplitude $t_{\rm exch}$,
reduces in the nonrelativistic limit to
$$
k^2 = - {\bf k}^2\ .
$$
The spinor factors $\bar u\,\gamma^\mu\,u$ and $\bar v\,\gamma^\mu\,v$ may be
evaluated very easily:
\begin{itemize}
\item For our particular choice for the normalization of the Dirac
spinors\index{Dirac spinor!---, normalization} $u$ and $v$, we have, in the
nonrelativistic limit,
\begin{eqnarray*}
\bar u(\tau_1)\,\gamma_0\,u(\sigma_1) &\equiv& u^\dagger(\tau_1)\,u(\sigma_1)
= \delta_{\tau_1\sigma_1}\ ,\\[1ex]
\bar v(\sigma_2)\,\gamma_0\,v(\tau_2) &\equiv& v^\dagger(\sigma_2)\,v(\tau_2)
= \delta_{\tau_2\sigma_2}\ .
\end{eqnarray*}
\item In the Dirac representation, the Dirac matrices\index{Dirac matrix!---,
Dirac representation} $\gamma^\mu \equiv
\{\gamma^0,\mbox{\boldmath$\gamma$}\}$ are explicitly given by
\begin{eqnarray*}
\gamma_0 &=& \left(\begin{array}{cr}1&0\\ 0&-1\end{array}\right)\ ,\\[1ex]
\mbox{\boldmath$\gamma$} &=& \left(\begin{array}{rc}0&
\mbox{\boldmath$\sigma$}\\ -\mbox{\boldmath$\sigma$}&0\end{array}\right)\ .
\end{eqnarray*}
Inserting these explicit representations of the Dirac matrices, we obtain, in
the nonrelativistic limit,
\begin{eqnarray*}
\bar u(\tau_1)\,\mbox{\boldmath$\gamma$}\,u(\sigma_1)
&\equiv& u^\dagger(\tau_1)\,\gamma_0\,\mbox{\boldmath$\gamma$}\,u(\sigma_1)
\\[1ex]
&=& u^\dagger(\tau_1)\left(\begin{array}{cr}1&0\\ 0&-1\end{array}\right)
\left(\begin{array}{rc}0&\mbox{\boldmath$\sigma$}\\
-\mbox{\boldmath$\sigma$}&0\end{array}\right)u(\sigma_1)\\[1ex]
&=& u^\dagger(\tau_1)\left(\begin{array}{cc}0&\mbox{\boldmath$\sigma$}\\
\mbox{\boldmath$\sigma$}&0\end{array}\right)u(\sigma_1)\\[1ex]
&=& \left({\chi_{\tau_1}}^\dagger,0\right)
\left(\begin{array}{cc}0&\mbox{\boldmath$\sigma$}\\
\mbox{\boldmath$\sigma$}&0\end{array}\right)\left(\begin{array}{c}
\chi_{\sigma_1}\\ 0\end{array}\right)\\[1ex]
&=& 0
\end{eqnarray*}
and, similarly,
$$
\bar v(\sigma_2)\,\mbox{\boldmath$\gamma$}\,v(\tau_2) = 0\ .
$$
\end{itemize}
Accordingly, the scattering amplitude $t_{\rm exch}$ of Eq.~(\ref{eq:t-rel})
reduces to
$$
t_{\rm exch} = \frac{e^2}{{\bf k}^2}
\,\delta_{\tau_1\sigma_1}\,\delta_{\tau_2\sigma_2}\ .
$$
\item The contribution of the pair annihilation\index{pair annihilation}
graph in Fig.~\ref{fig:elposscat-ann} to our scattering amplitude $t$ reads
$$
t_{\rm ann} = \frac{e^2}{(p_1 + p_2)^2}\,
\bar u(q_1,\tau_1)\,\gamma_\mu\,v(q_2,\tau_2)\,
\bar v(p_2,\sigma_2)\,\gamma^\mu\,u(p_1,\sigma_1)\ .
$$
Here, the total momentum $P$ of the system under consideration,
$$
P \equiv p_1 + p_2 = q_1 + q_2\ ,
$$
enters in the denominator of the scattering amplitude $t_{\rm ann}$. In that
case, however, the square of this total momentum,
\begin{eqnarray*}
P^2 &\equiv& (p_1 + p_2)^2\\[1ex]
&=& (E_{p_1} + E_{p_2})^2 - ({\bf p}_1 + {\bf p}_2)^2\ ,
\end{eqnarray*}
reduces in the nonrelativistic limit to
$$
P^2 \equiv (p_1 + p_2)^2 = (2\,m)^2\ .
$$
Thus, compared with the contribution to the scattering amplitude $t$ arising
from one-photon exchange, $t_{\rm exch}$, the contribution to the scattering
amplitude $t$ arising from pair annihilation, $t_{\rm ann}$, will be of the
order of
$$
\frac{{\bf k}^2}{m^2}\ .
$$
This observation indicates that the annihilation contribution $t_{\rm ann}$
represents, in any case, already some relativistic correction to the exchange
contribution $t_{\rm exch}$, for which there will be no room at all within a
purely nonrelativistic investigation and which, therefore, has to be
neglected for the present discussion.
\end{itemize}
We are unambiguously led to the conclusion that, in the nonrelativistic
limit, only the one-photon exchange\index{one-photon exchange} graph
contributes to the T-matrix element for elastic electron--positron
scattering:
$$
T_{\rm f\/i} = \frac{1}{(2\pi)^6}\,t
= \frac{1}{(2\pi)^6}\,\frac{e^2}{{\bf k}^2}
\,\delta_{\tau_1\sigma_1}\,\delta_{\tau_2\sigma_2}\ .
$$

According to step 3 of our procedure, the interaction potential $V({\bf x})$
is obtained as the Fourier transform of the T-matrix element $T_{\rm f\/i}$.
Since, at present, we are exclusively interested in the nonrelativistic
limit, we shall obtain in this way only the nonrelativistic (or static) part
$V_{\rm NR}({\bf x})$ of the potential:
\begin{eqnarray*}
V_{\rm NR}({\bf x})
&=& - (2\pi)^3\int d^3k\exp(-i\,{\bf k}\cdot{\bf x})\,T_{\rm f\/i}(k)
\\[1ex]
&=& - \frac{1}{(2\pi)^3}\int d^3k\exp(-i\,{\bf k}\cdot{\bf x})\,t
\\[1ex]
&=& - \frac{1}{(2\pi)^3}\,e^2\int d^3k\,
\frac{\exp(-i\,{\bf k}\cdot{\bf x})}{{\bf k}^2}\ .
\end{eqnarray*}

The result of the required integration may immediately be written down:
\begin{enumerate}
\item The integral is obviously invariant under rotations. Consequently, it
has to be some function $\Phi$ of the radial coordinate $r \equiv |{\bf x}|$
only:
$$
\int d^3k\,\frac{\exp(-i\,{\bf k}\cdot{\bf x})}{{\bf k}^2} = \Phi(r)\ .
$$
\item For dimensional reasons, this function $\Phi(r)$ has to be proportional
to the inverse of $r$:
$$
\Phi(r) \propto \dfrac{1}{r}\ .
$$
\end{enumerate}
These considerations justify the ansatz
$$
\frac{1}{(2\pi)^3}\int d^3k\,\frac{\exp(-i\,{\bf k}\cdot{\bf x})}{{\bf k}^2}
= \frac{A}{r}\ ,
$$
with some dimensionless constant $A$. We determine the constant $A$ by
applying the Laplacian $\Delta \equiv \mbox{\boldmath
$\nabla$}\cdot\mbox{\boldmath$\nabla$}$ to both sides of this ansatz:
\begin{itemize}
\item For the left-hand side, we find
\begin{eqnarray*}
\frac{1}{(2\pi)^3}\,\Delta\int d^3k\,
\frac{\exp(-i\,{\bf k}\cdot{\bf x})}{{\bf k}^2}
&=& - \frac{1}{(2\pi)^3}\int d^3k\exp(-i\,{\bf k}\cdot{\bf x})\\[1ex]
&=& - \delta^{(3)}({\bf x})\ .
\end{eqnarray*}
\item For the right-hand side, upon remembering the relation
$$
\Delta\frac{1}{r} = - 4\pi\,\delta^{(3)}({\bf x})\ ,
$$
we find
$$
A\,\Delta\frac{1}{r} = - 4\pi\,A\,\delta^{(3)}({\bf x})\ .
$$
\end{itemize}
By comparison, the dimensionless proportionality factor, $A$, is pinned down
to the value
$$
A = \frac{1}{4\pi} \ .
$$

With due satisfaction, we realize that, by following step by step our general
prescription given in our introductory remarks to this chapter, one is indeed
able to recover, from the nonrelativistic limit of the Born approximation to
the T-matrix element for (elastic) electron--positron scattering, the static
Coulomb potential of quantum electrodynamics:
$$
V_{\rm NR}^{\rm QED}(r) = -\dfrac{e^2}{4\pi\,r}\ ,
$$
or, with the usually employed definition
$$
\alpha_{\rm em} \equiv \frac{e^2}{4\pi}
$$
of the electromagnetic fine structure constant,\index{electromagnetic fine
structure constant}\index{fine structure constant!---,
electromagnetic}\index{electron--positron scattering|)}\index{quantum
electrodynamics|)}\index{QED|)}
$$
\fbox{\ $V_{\rm NR}^{\rm QED}(r) = - \dfrac{\alpha_{\rm em}}{r}$\ }\ .
$$

\section{Static potential in quantum
chromodynamics}\label{sec:pot-qcd}\index{quantum
chromodynamics|(}\index{QCD|(}

The overwhelming success in the case of quantum electrodynamics has
contributed to enhance our confidence in our prescription of extracting the
(perturbatively accessible part of an) effective interaction potential from
the relevant elastic-scattering problem. Hence, we do not hesitate to apply
this procedure also to the case of quantum chromodynamics.

The relevant situation for the determination of the potential which describes
the quark forces acting within mesons is the quark--antiquark scattering
$$
q_i(p_1,\sigma_1) + \bar q_j(p_2,\sigma_2) \rightarrow
q_k(q_1,\tau_1) + \bar q_\ell(q_2,\tau_2)\ ,
$$
where the indices $i,j,\dots = 1,2,3$ denote the colour degrees of freedom of
the involved quarks.

According to our brief but nevertheless comprehensive---not to say,
exhaustive---sketch of quantum chromodynamics given in Appendix~\ref{ch:qcd},
the coupling, with the interaction strength $g_{\rm s}$, of a quark $q$,
represented by the Dirac spinor field $q_i(x)$, to the gluon fields
$G^\mu_a(x)$, $a = 1,2,\dots,8$, is described by the interaction Lagrangian
$$
{\cal L}_{\rm I}^{\rm QCD}(x)
= g_{\rm s}\,\bar q_i(x)\,\gamma_\mu\,\frac{\lambda^a_{ij}}{2}\,q_j(x)\,
G^\mu_a(x)\ ,
$$
where $\lambda_a$, $a = 1,2,\dots,8$, are the eight Gell-Mann matrices; an
explicit representation of these matrices may be found in
Appendix~\ref{app:gellmannmatrices}. They serve to construct a fundamental
(three-dimensional) representation of the generators of the gauge group SU(3)
of quantum chromodynamics:
$$
T^a_{\rm fund} = \frac{\lambda_a}{2}\ .
$$

\begin{figure}[htb]
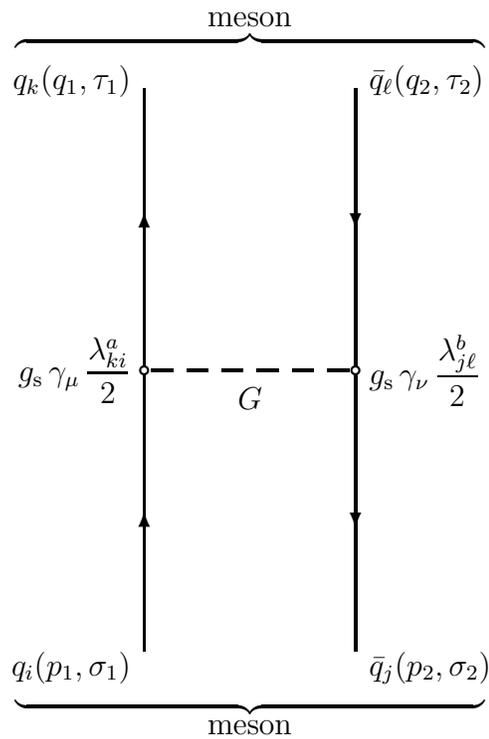

\normalsize
\begin{center}
$$
\underbrace{\overbrace{
\begin{array}{c}\\
\hspace{8.5ex}\qexchp\hspace{8.5ex}\\[2.5ex]
\end{array}}^{\mbox{meson}}}_{\mbox{meson}}
$$
\end{center}
\large
\caption{Quark--antiquark scattering, one-gluon exchange
graph.}\index{one-gluon exchange}\label{fig:qqbscat-ex}
\end{figure}

\begin{figure}[htb]
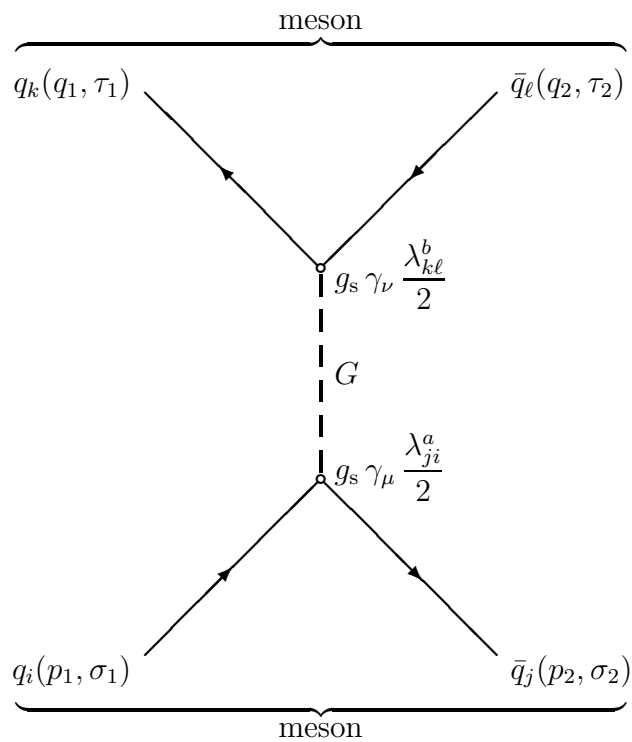

\normalsize
\begin{center}
$$
\underbrace{\overbrace{
\begin{array}{c}\\
\hspace{8.5ex}\qannp\hspace{8.5ex}\\[2.5ex]
\end{array}}^{\mbox{meson}}}_{\mbox{meson}}
$$
\end{center}
\large
\caption{Quark--antiquark scattering, pair annihilation
graph.}\index{pair annihilation}\label{fig:qqbscat-ann}
\end{figure}

Because of the (structural) similarity of the interaction Lagrangians ${\cal
L}_{\rm I}$ of quantum electrodynamics, ${\cal L}_{\rm I}^{\rm QED}$, and
quantum chromodynamics, ${\cal L}_{\rm I}^{\rm QCD}$, again only two Feynman
graphs potentially contribute, in lowest non-trivial order of the
perturbative loop expansion, to the scattering amplitude $T_{\rm f\/i}$ for
elastic quark--antiquark scattering:
\begin{itemize}
\item the exchange\index{one-gluon exchange} of a single gluon, $G_a$,
between quark and antiquark, as depicted in Fig.~\ref{fig:qqbscat-ex}; and
\item the ``annihilation''\index{pair annihilation} of the quark--antiquark
pair into a single gluon, $G_a$, followed by the subsequent creation of a
quark--antiquark pair by this single gluon, as depicted in
Fig.~\ref{fig:qqbscat-ann}.
\end{itemize}

\clearpage

There is absolutely no need to calculate the corresponding T-matrix elements
$T_{\rm f\/i}$ once more. Comparing the above interaction Lagrangians ${\cal
L}_{\rm I}$ of quantum electrodynamics, ${\cal L}_{\rm I}^{\rm QED}$, and
quantum chromodynamics, ${\cal L}_{\rm I}^{\rm QCD}$, we realize
that---loosely speaking---we may obtain the transition amplitudes required in
the present case from the ones computed in the previous section by simply
replacing in the latter the electric charge $e$ by the expression
$$
g_{\rm s}\,\frac{\lambda^a}{2}\ .
$$

However, we have to take into account that, according to the famous
confinement hypothesis, all the quarks inside a hadron form necessarily a
colour-singlet state. Consequently, we feel obliged to amend---which, as
there is no danger of confusion, we do without change of notations---the
transition amplitudes $T_{\rm f\/i}$ for the scattering of free particles by
the (normalized) meson colour wave functions
$$
\frac{1}{\sqrt{3}}\,\delta_{ij}\ ,
$$
which means, in fact, nothing else but an appropriate average over the colour
degrees of freedom.

With this proviso, the T-matrix element $T_{\rm f\/i}$ in question is found
as follows:
\begin{itemize}
\item The contribution of the one-gluon exchange\index{one-gluon exchange}
graph in Fig.~\ref{fig:qqbscat-ex} to our scattering amplitude $t$ reads
$$
t_{\rm exch} = - \frac{g_{\rm s}^2}{k^2}\,
\frac{\lambda^a_{ki}}{2}\,\frac{\lambda^a_{j\ell}}{2}\,
\bar u(q_1,\tau_1)\,\gamma_\mu\,u(p_1,\sigma_1)\,
\bar v(p_2,\sigma_2)\,\gamma^\mu\,v(q_2,\tau_2)\ ,
$$
where $k$ denotes again the involved momentum transfer,
$$
k \equiv p_1 - q_1 = q_2 - p_2\ ,
$$
or, after the announced multiplication by those meson colour wave functions,
\begin{eqnarray*}
t_{\rm exch} &=& - \frac{g_{\rm s}^2}{k^2}\,\frac{1}{\sqrt{3}}\,\delta_{ij}\,
\frac{1}{\sqrt{3}}\,\delta_{k\ell}\,\frac{\lambda^a_{ki}}{2}\,
\frac{\lambda^a_{j\ell}}{2}\\[1ex]
&\times& \bar u(q_1,\tau_1)\,\gamma_\mu\,u(p_1,\sigma_1)\,
\bar v(p_2,\sigma_2)\,\gamma^\mu\,v(q_2,\tau_2)\ .
\end{eqnarray*}
With one of the relations given in Appendix~\ref{app:gmm-tr}, the colour
factor stemming from the exchange graph yields
\begin{eqnarray*}
&{}& \underbrace{\frac{1}{\sqrt{3}}\,\sum_{i,j=1}^3\delta_{ij}\,
\frac{1}{\sqrt{3}}\,\sum_{k,\ell=1}^3\delta_{k\ell}}
_{\mbox{meson colour wave functions}} \times
\underbrace{\sum_{a=1}^{8}\frac{\lambda_{ki}^a}{2}\,
\frac{\lambda_{j\ell}^a}{2}}
_{\mbox{``(colour charge)${}^2$''$/g_{\rm s}^2$}}\\[1ex]
&{}& \qquad = \frac{1}{12}\,\sum_{i,k =1}^3\sum_{a=1}^8\lambda_{ki}^a\,
\lambda_{ik}^a\\[1ex]
&{}& \qquad = \frac{1}{12}\,\underbrace{\sum_{a=1}^8\mbox{Tr}
\left[(\lambda^a)^2\right]}_{\displaystyle 16}\\[1ex]
&{}& \qquad =\frac{4}{3}\ .
\end{eqnarray*}
Hence, the one-gluon exchange contribution $t_{\rm exch}$ to the scattering
amplitude $t$ is given by the expression
$$
t_{\rm exch} = - \frac{4}{3}\,\frac{g_{\rm s}^2}{k^2}\,
\bar u(q_1,\tau_1)\,\gamma_\mu\,u(p_1,\sigma_1)\,
\bar v(p_2,\sigma_2)\,\gamma^\mu\,v(q_2,\tau_2)\ ,
$$
which reduces in the nonrelativistic limit to
$$
t_{\rm exch} = \frac{4}{3}\,\frac{g_{\rm s}^2}{{\bf k}^2}
\,\delta_{\tau_1\sigma_1}\,\delta_{\tau_2\sigma_2}\ .
$$
\item The contribution of the pair annihilation\index{pair annihilation}
graph in Fig.~\ref{fig:qqbscat-ann} to our scattering amplitude $t$ reads
$$
t_{\rm ann}
= \frac{g_{\rm s}^2}{(p_1 + p_2)^2}\,
\frac{\lambda^a_{ji}}{2}\,\frac{\lambda^a_{k\ell}}{2}\,
\bar u(q_1,\tau_1)\,\gamma_\mu\,v(q_2,\tau_2)\,
\bar v(p_2,\sigma_2)\,\gamma^\mu\,u(p_1,\sigma_1)\ ,
$$
or, after the announced multiplication by those meson colour wave functions,
\begin{eqnarray*}
t_{\rm ann}
&=& \frac{g_{\rm s}^2}{(p_1 + p_2)^2}\,
\frac{1}{\sqrt{3}}\,\delta_{ij}\,
\frac{1}{\sqrt{3}}\,\delta_{k\ell}\,
\frac{\lambda^a_{ji}}{2}\,\frac{\lambda^a_{k\ell}}{2}\\[1ex]
&\times& \bar u(q_1,\tau_1)\,\gamma_\mu\,v(q_2,\tau_2)\,
\bar v(p_2,\sigma_2)\,\gamma^\mu\,u(p_1,\sigma_1)\ .
\end{eqnarray*}
However, this annihilation contribution $t_{\rm ann}$ vanishes identically:
\begin{eqnarray*}
\frac{1}{\sqrt{3}}\,\sum_{i,j=1}^3\delta_{ij}\,\frac{\lambda^a_{ji}}{2}
&\equiv& \frac{1}{2\sqrt{3}}\,\mbox{Tr}(\lambda^a)\\[1ex]
&=& 0\ .
\end{eqnarray*}
From the physical point of view, the interpretation of this, at first sight
slightly surprising, phenomenon is rather evident: the gluon, as a colour
octet, that is, as a particle which transforms according to the
eight-dimensional adjoint representation of SU(3)$_{\rm C}$, has no means to
couple to a colour singlet, like any bound state of quarks, without violating
thereby the conservation of colour demanded by the exact invariance of
quantum chromodynamics with respect to the colour gauge group SU(3)$_{\rm
C}$.
\end{itemize}

Collecting all our above findings, we may state that, in lowest order of the
perturbative loop expansion, the T-matrix element $T_{\rm f\/i}$ for elastic
quark--antiquark scattering within mesons receives only a contribution from
the one-gluon exchange\index{one-gluon exchange} graph:
$$
T_{\rm f\/i} = \frac{1}{(2\pi)^6}\,t
= \frac{1}{(2\pi)^6}\,\frac{4}{3}\,\frac{g_{\rm s}^2}{{\bf k}^2}
\,\delta_{\tau_1\sigma_1}\,\delta_{\tau_2\sigma_2}\ .
$$
Consequently, the quintessence of the present consideration is: we may recover
the (perturbatively accessible part of the) effective interaction
potential operative in quantum chromodynamics from its counterpart in the
case of quantum electrodynamics by simply replacing the square $e^2$ of the
electric charge $e$ by the factor
$$
\frac{4}{3}\,g_{\rm s}^2\ .
$$
Hence, the short-distance part of the static quark--antiquark potential,
arising from one-gluon exchange within mesons, is of Coulombic shape:
$$
V_{\rm NR}^{\rm QCD}(r) = - \dfrac{4}{3}\,\dfrac{g_{\rm s}^2}{4\pi\,r}\ ,
$$
or, with the usually employed definition
$$
\alpha_{\rm s} \equiv \frac{g_{\rm s}^2}{4\pi}
$$
of the strong fine structure constant,\index{strong fine structure
constant}\index{fine structure constant!---, strong}\index{quantum
chromodynamics|)}\index{QCD|)}
$$
\fbox{\ $V_{\rm NR}^{\rm QCD}(r)
= - \dfrac{4}{3}\,\dfrac{\alpha_{\rm s}}{r}$\ }\ .
$$

\section{Lorentz structure of an interquark
interaction}\label{sec:spinstruc}\index{spin structure|(}\index{Lorentz
structure|(}

At this stage, in order to seize hold of the nonperturbative contribution to
any effective potential, we embark on a rather general investigation.

Quite generally, the T-matrix element $T_{\rm f\/i}$ for the elastic
scattering of some generic fermion ${\cal F}$ and the corresponding
antifermion $\bar{\cal F}$, both of them of mass $m$,
$$
{\cal F}(p_1,\sigma_1) + \bar{\cal F}(p_2,\sigma_2) \rightarrow
{\cal F}(q_1,\tau_1) + \bar{\cal F}(q_2,\tau_2)\ ,
$$
is, apart from the overall normalization factors of the one-particle
wave functions, which we always split off by the definition
$$
T_{\rm f\/i} =:
\frac{1}{(2\pi)^6}\,\frac{m^2}{\sqrt{E_{p_1}\,E_{p_2}\,E_{q_1}\,E_{q_2}}}\,t
$$
of our scattering amplitude $t$, the product of
\begin{itemize}
\item two bilinears of Dirac spinors of the form $\bar
u(q_1,\tau_1)\,\Gamma_1\,u(p_1,\sigma_1)$ and $\bar
v(p_2,\sigma_2)\,\Gamma_2\,v(q_2,\tau_2)$, where $\Gamma_1$ and $\Gamma_2$
represent some (unspecified) Dirac matrices, and
\item some interaction kernel $K$ which, a priori, may depend on all four
external momenta $p_1,p_2,q_1,q_2$,
$$
K = K(p_1,p_2,q_1,q_2)\ ,
$$
only subject to the momentum conservation
$$
p_1 + p_2 = q_1 + q_2\ ,
$$
as expressed by that overall $\delta$ function multiplying this T-matrix
element $T_{\rm f\/i}$ in the standard decomposition of the S-matrix element
$S_{\rm f\/i}$.
\end{itemize}
Consequently, the most general ansatz for our scattering amplitude $t$ reads
$$
t =
\bar u(q_1,\tau_1)\,\Gamma_1\,u(p_1,\sigma_1)\,
\bar v(p_2,\sigma_2)\,\Gamma_2\,v(q_2,\tau_2)\,K\ ,
$$
with $K$ depending on any three independent linear combinations built from
the external momenta out of the set $\{p_1,p_2,q_1,q_2\}$.

We shall constrain the T-matrix element $T_{\rm f\/i}$ under consideration by
the following two, very reasonable assumptions:
\begin{enumerate}
\item The T-matrix element $T_{\rm f\/i}$ is invariant with respect to the
full set of transformations forming the (homogeneous) Lorentz group, that is,
invariant under
\begin{itemize}
\item proper orthochronous Lorentz transformations,
\item space reflection (``parity operation''), and
\item time reversal.
\end{itemize}
\item The interaction kernel $K$ entering in the T-matrix element $T_{\rm
f\/i}$ is a function of only the square $k^2$ of the involved momentum
transfer
$$
k \equiv p_1 - q_1 = q_2 - p_2\ ;
$$
that is,
$$
K = K(k^2)\ ,
$$
which reduces in the nonrelativistic limit to
$$
K = K(-{\bf k}^2)\ .
$$
\end{enumerate}

The most general form of the scattering amplitude $t$ consistent with the
requirements of the above assumptions is (see, e.g.,
Refs.~\cite{lucha91,gromes77,goldberger60})
\begin{equation}
t = \sum_{\Sigma = {\rm S},{\rm P},{\rm V},{\rm A},{\rm T}} t_\Sigma\ ,
\label{eq:t-ansatz}
\end{equation}
where any particular contribution $t_\Sigma$ is of the form
$$
t_\Sigma =
\bar u(q_1,\tau_1)\,\Gamma_\Sigma\,u(p_1,\sigma_1)\,
\bar v(p_2,\sigma_2)\,\Gamma_\Sigma\,v(q_2,\tau_2)\,K_\Sigma(k^2)
$$
and the sum extends over the five possible Lorentz structures
\begin{itemize}
\item scalar (S),
$$
\Gamma_\Sigma\otimes\Gamma_\Sigma = 1\otimes 1\ ,
$$
\item pseudoscalar (P),
$$
\Gamma_\Sigma\otimes\Gamma_\Sigma = \gamma_5\otimes\gamma^5\ ,
$$
\item vector (V),
$$
\Gamma_\Sigma\otimes\Gamma_\Sigma = \gamma_\mu\otimes\gamma^\mu\ ,
$$
\item axial vector (A),
$$
\Gamma_\Sigma\otimes\Gamma_\Sigma =
\gamma_\mu\,\gamma_5\otimes\gamma^\mu\,\gamma^5\ ,
$$
and
\item tensor (T),
$$
\Gamma_\Sigma\otimes\Gamma_\Sigma =
\frac{1}{2}\,\sigma_{\mu\nu}\otimes\sigma^{\mu\nu}\ .
$$
\end{itemize}
Here, we had to introduce the Dirac matrices\index{Dirac matrix}
\begin{eqnarray*}
\gamma_5 \equiv \gamma^5 &:=&
- \frac{i}{4!}\,\epsilon_{\mu\nu\rho\sigma}\,
\gamma^\mu\,\gamma^\nu\,\gamma^\rho\,\gamma^\sigma\\[1ex]
&=& -i\,\gamma^0\,\gamma^1\,\gamma^2\,\gamma^3\\
&=& i\,\gamma_0\,\gamma_1\,\gamma_2\,\gamma_3\ ,
\end{eqnarray*}
with the totally antisymmetric Levi--Civita symbol in four dimensions
$$
\epsilon_{\mu\nu\rho\sigma} = - \epsilon^{\mu\nu\rho\sigma}
$$
unambiguously fixed by demanding
$$
\epsilon_{0123} = 1\ ,
$$
as well as
$$
\sigma_{\mu\nu} := \frac{i}{2}\,[\gamma_\mu,\gamma_\nu]\ .
$$
In the Dirac representation, these Dirac matrices\index{Dirac matrix!---,
Dirac representation} are explicitly given by
$$
\gamma_5 = -\left(\begin{array}{cc}0&1\\ 1&0\end{array}\right)
$$
and
\begin{eqnarray*}
\sigma^{0i} &=& i\left(\begin{array}{cc}0&\sigma_i\\ \sigma_i&0\end{array}
\right)\ ,\quad i = 1,2,3\ ,\\[1ex]
\sigma^{ij} &=& \epsilon_{ijk}\left(\begin{array}{cc}\sigma_k&0\\0&\sigma_k
\end{array}\right)\ ,\quad i,j,k = 1,2,3\ .
\end{eqnarray*}

In the two preceding sections, that fermion--antifermion interaction was
basically mediated by the exchange of some vector boson, namely,
\begin{itemize}
\item the photon in the case of quantum electrodynamics or
\item the gluon in the case of quantum chromodynamics.
\end{itemize}
As a consequence of this, in both of these cases the effective interaction
was solely of vector Lorentz structure,
$$
\Gamma_1\otimes\Gamma_2 = \gamma_\mu\otimes\gamma^\mu\ ,
$$
or, in other words, in the decomposition (\ref{eq:t-ansatz}) the scattering
amplitude $t$ received exclusively a vector contribution. The interaction
kernel $K$ was given by
$$
K(k^2) = - \frac{\kappa}{k^2}
$$
reducing to
$$
K(-{\bf k}^2) = \frac{\kappa}{{\bf k}^2}\ ,
$$
where the effective coupling strength $\kappa$ stands
\begin{itemize}
\item in the case of quantum electrodynamics for
$$
\kappa = e^2\ ,
$$
\item in the case of quantum chromodynamics for
$$
\kappa = \frac{4}{3}\,g_{\rm s}^2\ .
$$
\end{itemize}
Fourier transformation then resulted in the static Coulomb potential
$$
V_{\rm NR}(r) = - \frac{\kappa}{4\pi\,r}\ .
$$

Needless to say, in general the various contributions entering in the
decomposition (\ref{eq:t-ansatz}) of our scattering amplitude $t$ will not
arise from the exchange of a single particle representing some fundamental
degree of freedom of the underlying quantum field theory. Rather, these terms
have to be interpreted as due to only an {\em effective\/} exchange of a
particle of the appropriate behaviour under Lorentz transformations.

We should be prepared to the fact that---in the course of evaluating below
the various terms $t_\Sigma$ contributing to the T-matrix element $T_{\rm
f\/i}$---we shall encounter expectation values of the Pauli matrices of the
form ${\chi_{\tau_1}}^\dagger\,\mbox{\boldmath$\sigma$}\,\chi_{\sigma_1}$ and
${\chi^c_{\sigma_2}}^\dagger\,\mbox{\boldmath$\sigma$}\,\chi^c_{\tau_2}$.
We shall cast the second of these expressions, which involves
two-component spinors $\chi^c_\sigma$ representing the spin degrees of freedom
of antifermions, defined by
$$
\chi^c_\sigma \equiv -i\,\sigma_2\,\chi^\ast_\sigma\ ,
$$
with the help of the identity
\begin{eqnarray*}
\sigma_2\,\mbox{\boldmath$\sigma$}\,\sigma_2
&=& - {\mbox{\boldmath$\sigma$}}^T\\
&=& - {\mbox{\boldmath$\sigma$}}^\ast
\end{eqnarray*}
into an equivalent form which involves only the two-component spinors
$\chi_\sigma$ pertaining to fermions:
\begin{eqnarray*}
{\chi^c_\sigma}^\dagger\,\mbox{\boldmath$\sigma$}\,\chi^c_\tau
&=& \left({\chi^c_\sigma}^\dagger\,\mbox{\boldmath$\sigma$}\,\chi^c_\tau
\right)^T\\[1ex]
&=& \left(\chi^T_\sigma\,\sigma_2\,\mbox{\boldmath$\sigma$}\,\sigma_2\,
\chi^\ast_\tau\right)^T\\[1ex]
&=& - \left(\chi^T_\sigma\,{\mbox{\boldmath$\sigma$}}^T\,\chi^\ast_\tau
\right)^T\\[1ex]
&=& - {\chi_\tau}^\dagger\,\mbox{\boldmath$\sigma$}\,\chi_\sigma\ .
\end{eqnarray*}
We shall find it convenient to abbreviate the expectation values of the Pauli
matrices $\mbox{\boldmath$\sigma$}$ by introducing the shorthand notation
\begin{equation}
\begin{array}{rcl}
\mbox{\boldmath$\sigma$}_1 &\equiv&
{\chi_{\tau_1}}^\dagger\,\mbox{\boldmath$\sigma$}\,\chi_{\sigma_1}\ ,\\[1ex]
\mbox{\boldmath$\sigma$}_2 &\equiv&
-{\chi^c_{\sigma_2}}^\dagger\,\mbox{\boldmath$\sigma$}\,\chi^c_{\tau_2}\\[1ex]
&=& {\chi_{\tau_2}}^\dagger\,\mbox{\boldmath$\sigma$}\,\chi_{\sigma_2}\ .
\label{eq:sigmaexpval}
\end{array}
\end{equation}

Now, what we really have to do when trying to follow the steps given in our
prescription for the derivation of that (perturbatively accessible part of
an) effective interaction potential from the underlying quantum field theory
by considering the relevant elastic-scattering problem may be phrased in
the following way:
\begin{enumerate}
\item Calculate the expectation values of the considered S operator,\index{S
operator} $S_{\rm f\/i}$, or T operator,\index{T operator} $T_{\rm f\/i}$,
with respect to those Fock-space\index{Fock space} states usually employed in
quantum field theory.
\item Look upon these matrix elements as the expectation values of that
interaction potential you are searching for, $V({\bf x})$, with respect to
the quantum-theoretical bound states and extract this interaction potential
by ``factorizing off'' all remnants of these bound states.
\end{enumerate}
In this and only this (!) sense one may adhere, when switching from the
scattering amplitude to the interaction potential, to the identifications of
the spin operators ${\bf S}_1$ and ${\bf S}_2$ of fermion and antifermion
with half of the expectation values $\mbox{\boldmath$\sigma$}_1$ and
$\mbox{\boldmath$\sigma$}_2$ of the Pauli matrices
$\mbox{\boldmath$\sigma$}$, respectively:
\begin{eqnarray*}
{\bf S}_1 &\mbox{``=''}& \frac{\mbox{\boldmath$\sigma$}_1}{2}\ ,\\[1ex]
{\bf S}_2 &\mbox{``=''}& \frac{\mbox{\boldmath$\sigma$}_2}{2}\ ,
\end{eqnarray*}
and therefore
$$
\mbox{\boldmath$\sigma$}_1\cdot\mbox{\boldmath$\sigma$}_2
\:\mbox{``=''}\: 4\,{\bf S}_1\cdot{\bf S}_2\ .
$$

Adopting an admittedly rather symbolical notation, we suppress in the
following any reference to both the momenta and the spin degrees of freedom
of the involved particles. Nevertheless, at every moment it should be clear
what's going on. With the above at one's disposal, the nonrelativistic
potentials $V^{(\Sigma)}_{\rm NR}({\bf x})$, $\Sigma = {\rm S},{\rm P},{\rm
V},{\rm A},{\rm T}$, are easily found:
\begin{description}
\item[Scalar:] For the scalar Lorentz structure, i.e.,
$$
\Gamma_\Sigma\otimes\Gamma_\Sigma = 1\otimes 1\ ,
$$
we find for our scattering amplitude $t$ in the nonrelativistic limit
\begin{eqnarray*}
t_{\rm S} &\equiv& \bar u\,u\,\bar v\,v\,K_{\rm S}(k^2)\\[1ex]
&=& -K_{\rm S}(-{\bf k}^2)\ .
\end{eqnarray*}
Upon Fourier transformation, the corresponding static interaction potential
$V_{\rm NR}^{({\rm S})}({\bf x})$ reads
$$
V_{\rm NR}^{({\rm S})}({\bf x}) = V_{\rm S}(r)\ ,
$$
where $V_{\rm S}(r)$ is defined by
$$
V_{\rm S}(r) \equiv \frac{1}{(2\pi)^3}\int d^3k
\exp(-i\,{\bf k}\cdot{\bf x})\,K_{\rm S}(-{\bf k}^2)\ .
$$
Accordingly, the scalar Lorentz structure yields a ``pure potential'' term.

\item[Pseudoscalar:] For the pseudoscalar Lorentz structure, i.e.,
$$
\Gamma_\Sigma\otimes\Gamma_\Sigma = \gamma_5\otimes\gamma^5\ ,
$$
we find for our scattering amplitude $t$ in the nonrelativistic limit
\begin{eqnarray*}
t_{\rm P}
&\equiv& \bar u\,\gamma_5\,u\,\bar v\,\gamma^5\,v\,K_{\rm P}(k^2)\\[1ex]
&=& u^\dagger\,\gamma_0\,\gamma_5\,u\,
v^\dagger\,\gamma_0\,\gamma^5\,v\,K_{\rm P}(k^2)\\[1ex]
&=& \left(\chi^\dagger,0\right)
\left(\begin{array}{rr}0&-1\\ 1&0\end{array}\right)
\left(\begin{array}{c}\chi\\ 0\end{array}\right)
\left(0,{\chi^c}^\dagger\right)
\left(\begin{array}{rr}0&-1\\ 1&0\end{array}\right)
\left(\begin{array}{c}0\\ \chi^c\end{array}\right)K_{\rm P}(-{\bf k}^2)\\[1ex]
&=& 0\ .
\end{eqnarray*}
Consequently, in the nonrelativistic limit, the contribution of this
pseudoscalar Lorentz structure vanishes:
$$
V_{\rm NR}^{({\rm P})}({\bf x}) = 0\ .
$$
\item[Vector:] For the vector Lorentz structure, i.e.,
$$
\Gamma_\Sigma\otimes\Gamma_\Sigma = \gamma_\mu\otimes\gamma^\mu\ ,
$$
the nonrelativistic limit of our scattering amplitude $t$ has, in fact,
already been calculated in Section~\ref{sec:pot-qed}:
\begin{eqnarray*}
t_{\rm V}
&\equiv& \bar u\,\gamma_\mu\,u\,\bar v\,\gamma^\mu\,v\,K_{\rm V}(k^2)\\[1ex]
&=& K_{\rm V}(-{\bf k}^2)\ .
\end{eqnarray*}
Upon Fourier transformation, the corresponding static interaction potential
$V_{\rm NR}^{({\rm V})}({\bf x})$ reads
$$
V_{\rm NR}^{({\rm V})}({\bf x}) = V_{\rm V}(r)\ ,
$$
where $V_{\rm V}(r)$ is defined by
$$
V_{\rm V}(r) \equiv - \frac{1}{(2\pi)^3}\int d^3k
\exp(-i\,{\bf k}\cdot{\bf x})\,K_{\rm V}(-{\bf k}^2)\ .
$$
Accordingly, the vector Lorentz structure yields, very similarly to its
scalar counterpart, a ``pure potential'' term.
\item[Axial vector:] For the axial vector Lorentz structure, i.e.,
$$
\Gamma_\Sigma\otimes\Gamma_\Sigma =
\gamma_\mu\,\gamma_5\otimes\gamma^\mu\,\gamma^5\ ,
$$
we find
\begin{itemize}
\item for $\mu = 0$
\begin{eqnarray*}
\bar u\,\gamma_0\,\gamma_5\,u\,\bar v\,\gamma^0\,\gamma^5\,v
&\equiv& u^\dagger\,\gamma_5\,u\,v^\dagger\,\gamma^5\,v\\[1ex]
&=& \left(\chi^\dagger,0\right)
\left(\begin{array}{rr}0&-1\\ -1&0\end{array}\right)
\left(\begin{array}{c}\chi\\ 0\end{array}\right)\\[1ex]
&\times& \left(0,{\chi^c}^\dagger\right)
\left(\begin{array}{rr}0&-1\\ -1&0\end{array}\right)
\left(\begin{array}{c}0\\ \chi^c\end{array}\right)\\[1ex]
&=& 0
\end{eqnarray*}
and
\item for $\mu = i$
\begin{eqnarray*}
\bar u\,\mbox{\boldmath$\gamma$}\,\gamma_5\,u\cdot
\bar v\,\mbox{\boldmath$\gamma$}\,\gamma^5\,v &\equiv&
u^\dagger\,\gamma_0\,\mbox{\boldmath$\gamma$}\,\gamma_5\,u\cdot
v^\dagger\,\gamma^0\,\mbox{\boldmath$\gamma$}\,\gamma^5\,v\\[1ex]
&=& \left(\chi^\dagger,0\right)\left(\begin{array}{rr}
-\mbox{\boldmath$\sigma$}&0\\ 0&-\mbox{\boldmath$\sigma$}\end{array}\right)
\left(\begin{array}{c}\chi\\ 0\end{array}\right)\\[1ex]
&\cdot& \left(0,{\chi^c}^\dagger\right)\left(\begin{array}{rr}
-\mbox{\boldmath$\sigma$}&0\\ 0&-\mbox{\boldmath$\sigma$}\end{array}\right)
\left(\begin{array}{c}0\\ \chi^c\end{array}\right)\\[1ex]
&=& \chi^\dagger\,\mbox{\boldmath$\sigma$}\,\chi\cdot
{\chi^c}^\dagger\,\mbox{\boldmath$\sigma$}\,\chi^c\\[1ex]
&\equiv& - \mbox{\boldmath$\sigma$}_1\cdot\mbox{\boldmath$\sigma$}_2\ ,
\end{eqnarray*}
\end{itemize}
and therefore for our scattering amplitude $t$ in the nonrelativistic limit
\begin{eqnarray*}
t_{\rm A} &\equiv& \bar u\,\gamma_\mu\,\gamma_5\,u\,
\bar v\,\gamma^\mu\,\gamma^5\,v\,K_{\rm A}(k^2)\\[1ex]
&=& \left[\bar u\,\gamma_0\,\gamma_5\,u\,
\bar v\,\gamma^0\,\gamma^5\,v -
\bar u\,\mbox{\boldmath$\gamma$}\,\gamma_5\,u\cdot
\bar v\,\mbox{\boldmath$\gamma$}\,\gamma^5\,v\right]K_{\rm A}(k^2)\\[1ex]
&=& \mbox{\boldmath$\sigma$}_1\cdot\mbox{\boldmath$\sigma$}_2\,
K_{\rm A}(-{\bf k}^2)\ .
\end{eqnarray*}
Upon Fourier transformation, the corresponding static interaction potential
$V_{\rm NR}^{({\rm A})}({\bf x})$ reads
$$
V_{\rm NR}^{({\rm A})}({\bf x}) = 4\,{\bf S}_1\cdot{\bf S}_2\,V_{\rm A}(r)\ ,
$$
where $V_{\rm A}(r)$ is defined by
$$
V_{\rm A}(r) \equiv - \frac{1}{(2\pi)^3}\int d^3k
\exp(-i\,{\bf k}\cdot{\bf x})\,K_{\rm A}(-{\bf k}^2)\ .
$$
Accordingly, the axial vector Lorentz structure entails an effective
spin--spin interaction.
\item[Tensor:] For the tensor Lorentz structure, i.e.,
$$
\Gamma_\Sigma\otimes\Gamma_\Sigma =
\frac{1}{2}\,\sigma_{\mu\nu}\otimes\sigma^{\mu\nu}\ ,
$$
we find
\begin{itemize}
\item for $\mu=0,\nu=i$
\begin{eqnarray*}
\bar u\,\sigma^{0i}\,u\,\bar v\,\sigma^{0i}\,v &\equiv&
u^\dagger\,\gamma_0\,\sigma^{0i}\,u\,
v^\dagger\,\gamma^0\,\sigma^{0i}\,v\\[1ex]
&=& i\left(\chi^\dagger,0\right)\left(\begin{array}{cc}0&\sigma_i\\
-\sigma_i&0\end{array}\right)\left(\begin{array}{c}\chi\\ 0\end{array}\right)
\\[1ex]
&\times& i\left(0,{\chi^c}^\dagger\right)\left(\begin{array}{cc}0&\sigma_i\\
-\sigma_i&0\end{array}\right)\left(\begin{array}{c}0\\
\chi^c\end{array}\right)\\[1ex]
&=& 0
\end{eqnarray*}
and
\item for $\mu=i,\nu=j$
\begin{eqnarray*}
\bar u\,\sigma^{ij}\,u\,\bar v\,\sigma^{ij}\,v &\equiv&
u^\dagger\,\gamma_0\,\sigma^{ij}\,u\,
v^\dagger\,\gamma^0\,\sigma^{ij}\,v\\[1ex]
&=& \epsilon_{ijk}\left(\chi^\dagger,0\right)\left(\begin{array}{cc}
\sigma_k&0\\ 0&-\sigma_k\end{array}\right)
\left(\begin{array}{c}\chi\\0\end{array}\right)\\[1ex]
&\times& \epsilon_{ij\ell}\left(0,{\chi^c}^\dagger\right)
\left(\begin{array}{cc}\sigma_\ell&0\\ 0&-\sigma_\ell\end{array}\right)
\left(\begin{array}{c}0\\ \chi^c\end{array}\right)\\[1ex]
&=& -2\,\chi^\dagger\,\sigma_k\,\chi\,{\chi^c}^\dagger\,\sigma_k\,\chi^c\\[1ex]
&\equiv& -2\,\chi^\dagger\,\mbox{\boldmath$\sigma$}\,\chi\cdot
{\chi^c}^\dagger\,\mbox{\boldmath$\sigma$}\,\chi^c\\[1ex]
&\equiv& 2\,\mbox{\boldmath$\sigma$}_1\cdot\mbox{\boldmath$\sigma$}_2\ ,
\end{eqnarray*}
\end{itemize}
and therefore for our scattering amplitude $t$ in the nonrelativistic limit
\begin{eqnarray*}
t_{\rm T} &\equiv& \frac{1}{2}\,
\bar u\,\sigma_{\mu\nu}\,u\,\bar v\,\sigma^{\mu\nu}\,v\,K_{\rm T}(k^2)\\[1ex]
&=& \left[\bar u\,\sigma_{0i}\,u\,\bar v\,\sigma^{0i}\,v + \frac{1}{2}\,
\bar u\,\sigma_{ij}\,u\,\bar v\,\sigma^{ij}\,v\right]K_{\rm T}(k^2)\\[1ex]
&=& \mbox{\boldmath$\sigma$}_1\cdot\mbox{\boldmath$\sigma$}_2\,
K_{\rm T}(-{\bf k}^2)\ .
\end{eqnarray*}
Upon Fourier transformation, the corresponding static interaction potential
$V_{\rm NR}^{({\rm T})}({\bf x})$ reads
$$
V_{\rm NR}^{({\rm T})}({\bf x}) = 4\,{\bf S}_1\cdot{\bf S}_2\,V_{\rm T}(r)\ ,
$$
where $V_{\rm T}(r)$ is defined by
$$
V_{\rm T}(r) \equiv - \frac{1}{(2\pi)^3}\int d^3k
\exp(-i\,{\bf k}\cdot{\bf x})\,K_{\rm T}(-{\bf k}^2)\ .
$$
Accordingly, the tensor Lorentz structure entails also an effective
spin--spin interaction.
\end{description}
Table~\ref{tab:lorst} summarizes our findings for the contributions of the
various possible Lorentz structures to the effective interaction potential in
the nonrelativistic limit.
\begin{table}[hbt]
\caption{Nonrelativistic interaction potential $V^{(\Sigma)}_{\rm NR}$
arising effectively from the various conceivable Lorentz structures
$\Gamma_\Sigma\otimes\Gamma_\Sigma$ of an arbitrary fermion--antifermion
interaction}\label{tab:lorst}\index{Lorentz structure}
$$
\begin{array}{lcc}\hline\hline\\[-1.5ex]
\mbox{Lorentz structure}&\Gamma_\Sigma\otimes\Gamma_\Sigma
&\mbox{static potential}\\[1ex] \hline\\[-1.5ex]
\mbox{scalar}&1\otimes 1&V_{\rm S}(r)\\[1ex]
\mbox{pseudoscalar}&\gamma_5\otimes\gamma^5&0\\[1ex]
\mbox{vector}&\gamma_\mu\otimes\gamma^\mu&V_{\rm V}(r)\\[1ex]
\mbox{axial vector}&\quad\gamma_\mu\,\gamma_5\otimes\gamma^\mu\,\gamma^5\quad&
4\,{\bf S}_1\cdot{\bf S}_2\,V_{\rm A}(r)\\[1ex]
\mbox{tensor}&\frac{1}{2}\,\sigma_{\mu\nu}\otimes\sigma^{\mu\nu}&
4\,{\bf S}_1\cdot{\bf S}_2\,V_{\rm T}(r)\\[1ex] \hline\hline
\end{array}
$$
\end{table}

The total spin ${\bf S}$ of the respective bound state under consideration is
clearly given by the sum of the spins ${\bf S}_1$ and ${\bf S}_2$ of its
constituents:
$$
{\bf S} \equiv {\bf S}_1 + {\bf S}_2\ .
$$
Upon squaring this relation,
\begin{eqnarray*}
{\bf S}^2 &=& ({\bf S}_1 + {\bf S}_2)^2\\[1ex]
&=& {\bf S}_1^2 + {\bf S}_2^2 + 2\,{\bf S}_1\cdot{\bf S}_2\ ,
\end{eqnarray*}
we may express the product ${\bf S}_1\cdot{\bf S}_2$ of the two
spins ${\bf S}_1$ and ${\bf S}_2$ in terms of the squares of ${\bf
S}_1$, ${\bf S}_2$, and ${\bf S}$,
$$
{\bf S}_1\cdot{\bf S}_2
= \dfrac{1}{2}\left({\bf S}^2 - {\bf S}_1^2 - {\bf S}_2^2\right)\ ,
$$
and, therefore, its expectation values by the corresponding expectation
values of ${\bf S}_1^2$, ${\bf S}_2^2$, and ${\bf S}^2$:
$$
\langle{\bf S}_1\cdot{\bf S}_2\rangle
= \dfrac{1}{2}\left(\langle{\bf S}^2\rangle - \langle{\bf S}_1^2\rangle
- \langle{\bf S}_2^2\rangle\right)\ .
$$
Accordingly, expressed in terms of the quantum numbers $S$, $S_1$, and $S_2$
of the spins ${\bf S}$, ${\bf S}_1$, and ${\bf S}_2$, respectively, the
expectation values $\langle{\bf S}_1\cdot{\bf S}_2\rangle$ of the product of
the spins ${\bf S}_1$, ${\bf S}_2$ of the bound-state constituents read
$$
\langle{\bf S}_1\cdot{\bf S}_2\rangle = \frac{1}{2}\left[S\,(S+1)
- S_1\,(S_1+1) - S_2\,(S_2+1)\right]\ .
$$
For fermionic constituents with spin
$$
S_1 = S_2 = \dfrac{1}{2}\ ,
$$
we have
$$
S_1\,(S_1+1) = S_2\,(S_2+1) = \dfrac{3}{4}
$$
and therefore
$$
\langle{\bf S}_1\cdot{\bf S}_2\rangle
= \frac{1}{2}\,S\,(S+1) - \frac{3}{4}\ .
$$
Moreover, for fermionic constituents with spin
$$
S_1 = S_2 = \dfrac{1}{2}\ ,
$$
the quantum number $S$ of the total spin ${\bf S}$ may accept precisely
either of two values:
\begin{itemize}
\item $S=0$, which corresponds to some spin singlet, like the pion or the
$\eta$ meson in the case of light quarks, or the $\eta_{\rm c}$ in the
charmonium system.
\item $S=1$, which corresponds to some spin triplet, like the $\rho$,
$\omega$, and $\phi$ mesons in the case of light quarks, or the ${\rm
J}/\psi$ in the charmonium system, or the $\Upsilon$ in the bottomonium
system.
\end{itemize}
This implies for the eigenvalues $S\,(S+1)$ of the square ${\bf S}^2$ of the
total spin ${\bf S}$:
$$
S\,(S+1) = \left\{\begin{array}{rl}
0&\quad\mbox{for spin singlets, i.e.,}\ S = 0\ ,\\[1ex]
2&\quad\mbox{for spin triplets, i.e.,}\ S = 1\ .\end{array}\right.
$$
Accordingly, the expectation values $\langle{\bf S}_1\cdot{\bf S}_2\rangle$
of the product of the spins ${\bf S}_1$, ${\bf S}_2$ of the bound-state
constituents are finally given by
\begin{equation}
\langle{\bf S}_1\cdot{\bf S}_2\rangle = \left\{\begin{array}{rl}
-\dfrac{3}{4}&\quad\mbox{for spin singlets, i.e.,}\ S = 0\ ,\\[2ex]
+\dfrac{1}{4}&\quad\mbox{for spin triplets, i.e.,}\ S = 1\ .
\end{array}\right.
\label{eq:s1s2}
\end{equation}
The first and simultaneously most important lesson to be learned from the
above is that, for these two possible values of the quantum number $S$ of the
total spin ${\bf S}$ of the bound state, the spin--spin interaction term will
contribute necessarily with opposite signs.

Collecting all previous results, the following picture emerges for the {\em
nonrelativistic limit\/} of the most general effective fermion--antifermion
interaction potential $V_{\rm NR}({\bf x})$:
\begin{itemize}
\item Both scalar and vector Lorentz structures lead to ``pure potential''
terms:
$$
V^{(\Sigma)}_{\rm NR}({\bf x}) = V_\Sigma(r) \quad\mbox{for}\quad
\Sigma = {\rm S},{\rm V}\ .
$$
\item The contribution of the pseudoscalar Lorentz structure
vanishes:\footnote{\normalsize\ In the next---i.e., first non-trivial---order
of the present nonrelativistic expansion, this pseudoscalar Lorentz structure
contributes to the spin--spin and tensor interaction terms.}
$$
V_{\rm NR}^{({\rm P})}({\bf x}) = 0\ .
$$
This circumstance provides, for instance, a very compelling reason for the
(relatively) weak binding of deuterium\index{deuterium} in nuclear physics:
The interaction between nucleons, that is, protons and neutrons, is generally
accepted to be dominated by one-pion exchange\index{one-pion exchange}. Since
the $\pi$ meson is a {\em pseudoscalar\/} meson, the Lorentz structure of its
coupling to the nucleons has to be also of pseudoscalar nature in order to
form an interaction Lagrangian which is a Lorentz scalar. This fact implies
that only the relativistic corrections arising from one-pion exchange can be
responsible for the binding of a proton and a neutron to the deuterium.
\item Both axial vector and tensor Lorentz structures contribute merely
to the spin--spin interaction term:
$$
V^{(\Sigma)}_{\rm NR}({\bf x}) = 4\,{\bf S}_1\cdot{\bf S}_2\,V_\Sigma(r)
\quad\mbox{for}\quad \Sigma = {\rm A},{\rm T}\ .
$$
As a consequence of Eq.~(\ref{eq:s1s2}), in this case we will obtain a
binding force between fermion and antifermion only for either of the above
two possible values, $S=0$ and $S=1$, of the quantum number $S$ of the total
spin ${\bf S}$ of the bound state.
\end{itemize}

Hoping that the empirically observed hadron spectrum will provide some
restrictions on the allowed effective quark--antiquark interaction, we now
confront the above picture of general findings with experiment:
\begin{itemize}
\item Already the mere existence of strongly bound mesons forbids the
pseudoscalar Lorentz structure to play any significant r\^{o}le within some
phenomenologically acceptable quark--antiquark interaction potential.
\item The existence of both pseudoscalar mesons (like $\pi$, $\eta$, and
$\eta'$) {\em and\/} vector mesons (like $\rho$, $\omega$, and $\phi$), all
of which are bound states of a quark--antiquark pair with vanishing orbital
angular momentum, implies that the actual quark--antiquark forces must be
described by an interaction potential which yields binding for $S=0$ as well
as $S=1$. Obviously, this fact rules out both the axial vector and tensor
Lorentz structures as the predominant contribution to any realistic
quark--antiquark interaction potential.
\end{itemize}
In other words, the theoretically predicted particle spectra would look very
different from the experimentally measured ones if the dominant terms in the
effective quark--antiquark interaction potential would not be just some
linear combination of vector and scalar Lorentz structure. Therefore, our
conclusion has to be:
\begin{center}
\begin{tabular}{|c|}\hline\\[-2ex]
{\bf The Lorentz structure of the quark--antiquark interaction}\\
{\bf is dominated by the vector $\gamma_\mu \otimes \gamma^\mu$ and/or the
scalar $1 \otimes 1$,}\\
{\bf both of which lead in the nonrelativistic limit to so-called}\\
{\bf pure potential terms. Thus the static interaction potential}\\
{\bf $V_{\rm NR}(r)$ must be the sum of merely the contributions of the}\\
{\bf vector---$V_{\rm V}(r)$---and the scalar---$V_{\rm S}(r)$---Lorentz
structures:}\\[1ex]
{\bf $V_{\rm NR}(r) = V_{\rm V}(r) + V_{\rm S}(r)\ $.}\\[1ex]\hline
\end{tabular}
\end{center}
\index{nonrelativistic limit|)}\index{spin structure|)}\index{Lorentz
structure|)}

\chapter{Relativistic Corrections}\label{ch:breitfermi}\index{relativistic
corrections|(}\index{corrections!---, relativistic|(}

Beyond doubt, the next logical step must be to improve the up-to-now entirely
nonrelativistic formalism by taking into account all relativistic
corrections. In principle, one encounters no particular difficulties when
trying to take into account (at least, at some formal level) the complete set
of relativistic corrections to the effective interaction potential
\cite{lucha91rel,lucha92com}.

For the moment, however, we intend to be somewhat more modest, and this even
in two respects:
\begin{enumerate}
\item We shall calculate these relativistic corrections only up to second
order in the absolute value $v \equiv |{\bf v}|$ of the generic relative
velocities
$$
{\bf v} = \frac{{\bf p}}{E_p}
$$
of the bound-state constituents, that is, only up to order
$$
v^2 = \frac{{\bf p}^2}{E_p^2}\ ,
$$
which, since up to this order the relativistic kinetic energy
$$
E_p = \sqrt{{\bf p}^2 + m^2}
$$
may be approximated at this place by
$$
E_p \simeq m\ ,
$$
is equivalent to
$$
v^2 \simeq \frac{{\bf p}^2}{m^2}\ .
$$
\item We shall consider only the {\em spin-dependent\/} contributions to
these relativistic corrections. These spin-dependent interactions control the
fine and hyperfine level splittings of the bound-state spectra. The {\em
spin-independent\/} interactions may be obtained, with slightly more effort,
along similar lines \cite{gromes77}.
\end{enumerate}

\section{Spin-dependent corrections}\index{spin-dependent
corrections|(}\index{corrections!---, spin-dependent|(}

We shall be interested in all the spin-dependent relativistic corrections to
the static interaction potential $V_{\rm NR}({\bf x})$ up to order
$$
\frac{{\bf p}^2}{m^2}\ .
$$
Therefore, we must focus our attention to those terms in the transition
amplitude $T_{\rm f\/i}$ which involve expectation values of the Pauli
matrices\index{Pauli matrices} $\mbox{\boldmath$\sigma$}$. One may simplify
this task considerably by the following observation: we are entitled to
approximate the relativistic kinetic energy
$$
E_p = \sqrt{{\bf p}^2 + m^2}
$$
by the lowest-order term
$$
E_p \simeq m
$$
of its nonrelativistic expansion in two places, namely,
\begin{enumerate}
\item in the Dirac spinors\index{Dirac spinor} $u(p,\sigma)$ and $v(p,\sigma)$,
$$
\begin{array}{rcl}
u(p,\sigma) &=& \sqrt{\dfrac{E_p+m}{2\,m}}\left(\begin{array}{c}1\\[1ex]
\dfrac{\mbox{\boldmath$\sigma$}\cdot{\bf p}}{E_p+m}
\end{array}\right)\chi_\sigma\ ,\\
&&\\[-1ex]
v(p,\sigma) &=& \sqrt{\dfrac{E_p+m}{2\,m}}\left(\begin{array}{c}
\dfrac{\mbox{\boldmath$\sigma$}\cdot{\bf p}}{E_p+m}\\[2ex]
1\end{array}\right)\chi^c_\sigma\ ,\quad
\chi^c_\sigma \equiv -i\,\sigma_2\,\chi^\ast_\sigma\ ,
\end{array}
$$
which then (only for the purpose of the present analysis!) become
\begin{equation}
\begin{array}{rcl}
u(p,\sigma) &=& \left(\begin{array}{c}1\\[1ex]
\dfrac{\mbox{\boldmath$\sigma$}\cdot{\bf p}}{2\,m}
\end{array}\right)\chi_\sigma\ ,\\
&&\\[-1ex]
v(p,\sigma) &=& \left(\begin{array}{c}
\dfrac{\mbox{\boldmath$\sigma$}\cdot{\bf p}}{2\,m}\\[2ex]
1\end{array}\right)\chi^c_\sigma\ ;
\end{array}
\label{eq:spinor-relcorr}
\end{equation}
\item in that general relationship between the T-matrix element $T_{\rm f\/i}$
and our scattering amplitude $t$,
$$
T_{\rm f\/i} =:
\frac{1}{(2\pi)^6}\,\frac{m^2}{\sqrt{E_{p_1}\,E_{p_2}\,E_{q_1}\,E_{q_2}}}\,t\ ,
$$
which then becomes
$$
T_{\rm f\/i} = \frac{1}{(2\pi)^6}\,t\ .
$$
\end{enumerate}

In the course of calculating the scattering amplitude $t$, we may take
advantage of two trivial simplifications:
\begin{enumerate}
\item In order to get the interaction potential $V({\bf x})$, we have to
consider the scattering amplitude $t$ only in the center-of-momentum frame,
defined by the vanishing of the total momenta ${\bf P}_{\rm i}$ and ${\bf
P}_{\rm f}$ of initial state ${\rm i}$ and final state ${\rm f}$,
respectively:
$$
{\bf P}_{\rm i} \equiv {\bf p}_1 + {\bf p}_2 = {\bf q}_1 + {\bf q}_2
\equiv {\bf P}_{\rm f} = 0\ .
$$
Consequently, this scattering amplitude $t$ will depend only on the involved
momentum transfer
$$
{\bf k} \equiv {\bf p}_1 - {\bf q}_1 = {\bf q}_2 - {\bf p}_2
$$
and on the relative momentum
$$
{\bf p} \equiv {\bf p}_1 = -{\bf p}_2\ .
$$
\item The Pauli matrices\index{Pauli matrices} $\mbox{\boldmath$\sigma$}
\equiv \{\sigma_i, i = 1,2,3\}$ satisfy both
\begin{itemize}
\item the commutation relations
$$
[\sigma_i,\sigma_j] = 2\,i\,\epsilon_{ijk}\,\sigma_k
$$
and
\item the anticommutation relations
$$
\{\sigma_i,\sigma_j\} = 2\,\delta_{ij}\ .
$$
\end{itemize}
Adding up these two relations, the product $\sigma_i\,\sigma_j$ of any two
Pauli matrices is given by
\begin{eqnarray*}
\sigma_i\,\sigma_j &=&
\frac{1}{2}\left([\sigma_i,\sigma_j] + \{\sigma_i,\sigma_j\}\right)\\[1ex]
&=& \delta_{ij} + i\,\epsilon_{ijk}\,\sigma_k\ .
\end{eqnarray*}
By application of this relation, any product of (two or more) Pauli
matrices\index{Pauli matrices} may be reduced to an expression which involves
no more than at most {\em one\/} Pauli matrix.
\end{enumerate}

Performing the Fourier transformation as demanded by that general
prescription briefly sketched in our introductory remarks to
Chapter~\ref{ch:potential}, the effective interaction potential $V({\bf x})$
is derived from $t$ according to
\begin{eqnarray*}
V({\bf x})
&=& - (2\pi)^3\int d^3k\exp(-i\,{\bf k}\cdot{\bf x})\,T_{\rm f\/i}(k)\\[1ex]
&=& - \frac{1}{(2\pi)^3}\int d^3k\exp(-i\,{\bf k}\cdot{\bf x})\,t\ .
\end{eqnarray*}

As the central result of the intended inclusion of all spin-dependent
relativistic corrections up to order
$$
\frac{{\bf p}^2}{m^2}\ ,
$$
we shall finally end up with the generalization to arbitrary interaction
potentials of the well-known {\em Breit--Fermi
Hamiltonian}\index{Breit--Fermi Hamiltonian}, of the standard form
$$
H = m_1 + m_2 + \dfrac{{\bf p}^2}{2\,\mu} + V({\bf x})\ ,
$$
with $\mu$ the reduced mass of the particular two-particle quantum system
under consideration,
$$
\mu \equiv \frac{m_1\,m_2}{m_1 + m_2}\ .
$$
Here, the interaction potential $V({\bf x})$ will encompass, in addition to
the nonrelativistic contribution $V_{\rm NR}({\bf x})$, also all
spin-dependent relativistic corrections $V_{\rm spin}({\bf x})$:
$$
V({\bf x}) = V_{\rm NR}({\bf x}) + V_{\rm spin}({\bf x})\ .
$$
The set of spin-dependent relativistic corrections $V_{\rm spin}({\bf x})$
will turn out to consist, in general, of some
\begin{itemize}
\item spin--orbit interaction term $H_{\rm LS}$,
\item spin--spin interaction term $H_{\rm SS}$, and
\item tensor interaction term $H_{\rm T}$;
\end{itemize}
that is, this spin-dependent part $V_{\rm spin}$ of the interaction potential
$V({\bf x})$ will read
$$
V_{\rm spin} = H_{\rm LS} + H_{\rm SS} + H_{\rm T}\ .
$$

Bearing in mind the outcome of our analysis of the possible Lorentz structure
of the effective interaction in a quark--antiquark bound state as performed
in Section~\ref{sec:spinstruc}, we will treat below only the case of vector
and scalar Lorentz structure of the static interaction potential $V_{\rm
NR}(r)$:
$$
V_{\rm NR}(r) = V_{\rm V}(r) + V_{\rm S}(r)\ .
$$
Furthermore, for the sake of simplicity, we will present in the following all
the necessary derivations in detail only for the special case of equal masses
of the bound-state constituents: $$m_1 = m_2 = m\ .$$

\section{Interaction with vector Lorentz structure}\label{sec:relcor-vec}

In the case of an interaction with vector Lorentz structure, that is, for
$$
\Gamma_\Sigma \otimes \Gamma_\Sigma = \gamma_\mu \otimes \gamma^\mu\ ,
$$
the scattering amplitude (\ref{eq:t-ansatz}) assumes the form
$$
t_{\rm V} \equiv \bar u(q_1,\tau_1)\,\gamma_\mu\,u(p_1,\sigma_1)\,
\bar v(p_2,\sigma_2)\,\gamma^\mu\,v(q_2,\tau_2)\,K_{\rm V}(k^2)\ .
$$

Upon inserting the Dirac spinors in Dirac representation,
Eq.~(\ref{eq:spinor-relcorr}), and suppressing, for the sake of notational
simplicity, any reference to the spin degrees of freedom, we find for this
scattering amplitude
\begin{eqnarray*}
t_{\rm V}
&\equiv& \bar u(q_1)\,\gamma_\mu\,u(p_1)\,\bar v(p_2)\,\gamma^\mu\,v(q_2)\,
K_{\rm V}(k^2)\\[1ex]
&\equiv& \left[u^\dagger(q_1)\,u(p_1)\,v^\dagger(p_2)\,v(q_2)\right.\\[1ex]
&-& \left.u^\dagger(q_1)\,\gamma_0\,\mbox{\boldmath$\gamma$}\,u(p_1)\cdot
v^\dagger(p_2)\,\gamma_0\,\mbox{\boldmath$\gamma$}\,v(q_2)\right]
K_{\rm V}(k^2)\\[1ex]
&=& \left[\chi^\dagger
\left(1,\frac{\mbox{\boldmath$\sigma$}\cdot{\bf q}_1}{2\,m}\right)
\left(\begin{array}{c}1\\[1ex]
\dfrac{\mbox{\boldmath$\sigma$}\cdot{\bf p}_1}{2\,m}\end{array}\right)\chi\,
{\chi^c}^\dagger
\left(\frac{\mbox{\boldmath$\sigma$}\cdot{\bf p}_2}{2\,m},1\right)
\left(\begin{array}{c}\dfrac{\mbox{\boldmath$\sigma$}\cdot{\bf q}_2}{2\,m}
\\[2ex]1\end{array}\right)\chi^c\right.\\[1ex]
&-& \chi^\dagger
\left(1,\frac{\mbox{\boldmath$\sigma$}\cdot{\bf q}_1}{2\,m}\right)
\left(\begin{array}{cc}0&\mbox{\boldmath$\sigma$}\\
\mbox{\boldmath$\sigma$}&0\end{array}\right)
\left(\begin{array}{c}1\\[1ex]
\dfrac{\mbox{\boldmath$\sigma$}\cdot{\bf p}_1}{2\,m}\end{array}\right)\chi
\\[1ex]
&\cdot& \left.{\chi^c}^\dagger
\left(\frac{\mbox{\boldmath$\sigma$}\cdot{\bf p}_2}{2\,m},1\right)
\left(\begin{array}{cc}0&\mbox{\boldmath$\sigma$}\\
\mbox{\boldmath$\sigma$}&0\end{array}\right)
\left(\begin{array}{c}\dfrac{\mbox{\boldmath$\sigma$}\cdot{\bf q}_2}{2\,m}
\\[2ex]1\end{array}\right)\chi^c\right]K_{\rm V}(-{\bf k}^2)\\[1ex]
&=& \left\{\left[\chi^\dagger\,\chi + \frac{1}{4\,m^2}\,\chi^\dagger\,
(\mbox{\boldmath$\sigma$}\cdot{\bf q}_1)\,
(\mbox{\boldmath$\sigma$}\cdot{\bf p}_1)\,\chi\right]\right.\\[1ex]
&\times& \left.\left[{\chi^c}^\dagger\,\chi^c
+ \frac{1}{4\,m^2}\,{\chi^c}^\dagger\,
(\mbox{\boldmath$\sigma$}\cdot{\bf p}_2)\,
(\mbox{\boldmath$\sigma$}\cdot{\bf q}_2)\,\chi^c\right]\right.\\[1ex]
&-& \frac{1}{4\,m^2}\left[\chi^\dagger\,\mbox{\boldmath$\sigma$}\,
(\mbox{\boldmath$\sigma$}\cdot{\bf p}_1)\,\chi
+ \chi^\dagger\,(\mbox{\boldmath$\sigma$}\cdot{\bf q}_1)\,
\mbox{\boldmath$\sigma$}\,\chi\right]\\[1ex]
&\cdot& \left.\left[{\chi^c}^\dagger\,(\mbox{\boldmath$\sigma$}\cdot{\bf p}_2)
\,\mbox{\boldmath$\sigma$}\,\chi^c
+ {\chi^c}^\dagger\,\mbox{\boldmath$\sigma$}\,
(\mbox{\boldmath$\sigma$}\cdot{\bf q}_2)\,\chi^c\right]
\dfrac{}{}\!\!\right\}K_{\rm V}(-{\bf k}^2)\\[1ex]
&=& \left\{1 + \frac{1}{4\,m^2}\left\{\chi^\dagger\,
(\mbox{\boldmath$\sigma$}\cdot{\bf q}_1)\,
(\mbox{\boldmath$\sigma$}\cdot{\bf p}_1)\,\chi
+ {\chi^c}^\dagger\,(\mbox{\boldmath$\sigma$}\cdot{\bf p}_2)\,
(\mbox{\boldmath$\sigma$}\cdot{\bf q}_2)\,\chi^c\right.\right.\\[1ex]
&-& \chi^\dagger\,[\mbox{\boldmath$\sigma$}\,
(\mbox{\boldmath$\sigma$}\cdot{\bf p}_1) +
(\mbox{\boldmath$\sigma$}\cdot{\bf q}_1)\,\mbox{\boldmath$\sigma$}]\,\chi
\\[1ex]
&\cdot& \left.\left.{\chi^c}^\dagger\,
[(\mbox{\boldmath$\sigma$}\cdot{\bf p}_2)\,\mbox{\boldmath$\sigma$} +
\mbox{\boldmath$\sigma$}\,(\mbox{\boldmath$\sigma$}\cdot{\bf q}_2)]\,\chi^c
\right\}\dfrac{}{}\!\!\right\}K_{\rm V}(-{\bf k}^2)\ .
\end{eqnarray*}
Dropping all contributions to spin-independent relativistic corrections at
the very instant they show up, and recalling the abbreviations
(\ref{eq:sigmaexpval}), this scattering amplitude simplifies, in the
center-of-momentum frame, to
\begin{eqnarray*}
t_{\rm V}
&=& \left\{1 + \frac{1}{4\,m^2}\,\{i\,\epsilon_{ijk}\,p_i\,k_j\,\sigma_{1k}
+ i\,\epsilon_{ijk}\,p_i\,k_j\,\sigma_{2k}\right.\\[1ex]
&+& \left.\frac{}{}\![(2\,p - k)_i + i\,\epsilon_{ijk}\,k_j\,\sigma_{1k}]\,
[(2\,p - k)_i + i\,\epsilon_{i\ell m}\,k_\ell\,\sigma_{2m}]\}\right\}
K_{\rm V}(-{\bf k}^2)\\[1ex]
&=& \left\{1 + \frac{1}{4\,m^2}\,[3\,i\,\epsilon_{ijk}\,p_i\,k_j\,(\sigma_{1k}
+ \sigma_{2k})\right.\\[1ex]
&{}&\quad\quad\; - \left.
(\mbox{\boldmath$\sigma$}_1\cdot\mbox{\boldmath$\sigma$}_2)\,{\bf k}^2
+ (\mbox{\boldmath$\sigma$}_1\cdot{\bf k})\,
(\mbox{\boldmath$\sigma$}_2\cdot{\bf k})]\dfrac{}{}\!\right\}
K_{\rm V}(-{\bf k}^2)\ .
\end{eqnarray*}

Fourier transformation appears to be the appropriate tool to obtain the
resulting effective interaction potential:
\begin{itemize}
\item Already from the very beginning, all interaction kernels $K_\Sigma$
have been (implicitly) assumed to depend, in the nonrelativistic limit, only
on the modulus $|{\bf k}|$ of the momentum transfer ${\bf k}$. Therefore, all
the corresponding static interaction potentials $V^{(\Sigma)}_{\rm NR}({\bf
x})$ have to be spherically symmetric:
$$
V_{\rm V}(r) \equiv - \frac{1}{(2\pi)^3}\int d^3k
\exp(-i\,{\bf k}\cdot{\bf x})\,K_{\rm V}(-{\bf k}^2)\ .
$$
\item Denoting the first and second derivatives of any static interaction
potential with respect to the radial coordinate $r\equiv |{\bf x}|$ by
primes, one finds
\begin{eqnarray*}
\frac{1}{(2\pi)^3}\int d^3k\,k_j\exp(-i\,{\bf k}\cdot{\bf x})\,
K_{\rm V}(-{\bf k}^2)
&=& - i\,\nabla_j V_{\rm V}(r)\\[1ex]
&=& - i\,\frac{x_j}{r}\,V'_{\rm V}(r)\ ,
\end{eqnarray*}
$$
\frac{1}{(2\pi)^3}\int d^3k\,{\bf k}^2\exp(-i\,{\bf k}\cdot{\bf x})\,
K_{\rm V}(-{\bf k}^2) = \Delta V_{\rm V}(r)\ ,
$$
and, with the help of an identity proven in Appendix~\ref{app:nabla2},
\begin{eqnarray*}
&{}& \frac{1}{(2\pi)^3}\int d^3k\,k_i\,k_j\exp(-i\,{\bf k}\cdot{\bf x})\,
K_{\rm V}(-{\bf k}^2)\\[1ex]
&{}&\qquad = \nabla_i\nabla_j V_{\rm V}(r)\\[1ex]
&{}&\qquad = \left(\frac{x_i\,x_j}{r^2} - \frac{1}{3}\,\delta_{ij}\right)
\left[V''_{\rm V}(r) - \frac{1}{r}\,V'_{\rm V}(r)\right]
+ \frac{1}{3}\,\delta_{ij}\,\Delta V_{\rm V}(r)\ .
\end{eqnarray*}
\end{itemize}
Consequently, the spin-dependent relativistic corrections $V_{\rm spin}^{\rm
vector}$ for the case of a vector Lorentz structure of the effective
fermion--antifermion interaction become
\begin{eqnarray*}
V_{\rm spin}^{\rm vector}
&=& \frac{3}{2\,m^2\,r}\,({\bf x}\times{\bf p})\cdot{\bf S}\,
V_{\rm V}'(r)\\[1ex]
&+& \frac{2}{3\,m^2}\,{\bf S}_1\cdot{\bf S}_2\,\Delta V_{\rm V}(r)\\[1ex]
&+& \frac{1}{m^2}
\left[\frac{({\bf S}_1\cdot{\bf x})\,({\bf S}_2\cdot{\bf x})}{r^2}
- \frac{1}{3}\,{\bf S}_1\cdot{\bf S}_2\right]
\left[\frac{1}{r}\,V_{\rm V}'(r) - V_{\rm V}''(r)\right]\ .
\end{eqnarray*}

Herein, it is straightforward to identify, in full accordance with the
previously announced decomposition of the spin-dependent relativistic
corrections $V_{\rm spin}$, when specified to the case of vector Lorentz
structure,
$$
V_{\rm spin}^{\rm vector} =
H_{\rm LS}^{\rm vector} + H_{\rm SS}^{\rm vector} + H_{\rm T}^{\rm vector}\ ,
$$
\begin{itemize}
\item the spin--orbit term\index{spin--orbit term}
\begin{eqnarray*}
H_{\rm LS}^{\rm vector}
&=& \frac{3}{2\,m^2\,r}\,({\bf x}\times{\bf p})\cdot{\bf S}\,
V_{\rm V}'(r)\\[1ex]
&\equiv& \frac{3}{2\,m^2\,r}\,{\bf L}\cdot{\bf S}\,V_{\rm V}'(r)\ ,
\end{eqnarray*}
with the relative orbital angular momentum
$$
{\bf L} \equiv {\bf x}\times{\bf p}
$$
of the bound-state constituents;
\item the spin--spin term\index{spin--spin term}
$$
H_{\rm SS}^{\rm vector}
= \frac{2}{3\,m^2}\,{\bf S}_1\cdot{\bf S}_2\,\Delta V_{\rm V}(r)\ ;
$$
and
\item the tensor term\index{tensor term}
\begin{eqnarray*}
H_{\rm T}^{\rm vector} &=& \frac{1}{m^2}
\left[\frac{({\bf S}_1\cdot{\bf x})\,({\bf S}_2\cdot{\bf x})}{r^2}
- \frac{1}{3}\,{\bf S}_1\cdot{\bf S}_2\right]
\left[\frac{1}{r}\,V_{\rm V}'(r) - V_{\rm V}''(r)\right]\\[1ex]
&=& \frac{1}{12\,m^2}\,S_{12}
\left[\frac{1}{r}\,V_{\rm V}'(r) - V_{\rm V}''(r)\right]\ ,
\end{eqnarray*}
with the shorthand notation
$$
S_{12} \equiv 12
\left[\frac{({\bf S}_1\cdot{\bf x})\,({\bf S}_2\cdot{\bf x})}{r^2}
- \frac{1}{3}\,{\bf S}_1\cdot{\bf S}_2\right]
$$
for the spin-dependent factor.
\end{itemize}

\section{Interaction with scalar Lorentz structure}\label{sec:relcor-scal}

In the case of an interaction with scalar Lorentz structure, that is, for
$$
\Gamma_\Sigma \otimes \Gamma_\Sigma = 1 \otimes 1\ ,
$$
the scattering amplitude (\ref{eq:t-ansatz}) assumes the form
$$
t_{\rm S} \equiv \bar u(q_1,\tau_1)\,u(p_1,\sigma_1)\,
\bar v(p_2,\sigma_2)\,v(q_2,\tau_2)\,K_{\rm S}(k^2)\ .
$$

Upon inserting the Dirac spinors in Dirac representation,
Eq.~(\ref{eq:spinor-relcorr}), and suppressing, for the sake of notational
simplicity, any reference to the spin degrees of freedom, we find for this
scattering amplitude
\begin{eqnarray*}
t_{\rm S}
&\equiv& \bar u(q_1)\,u(p_1)\,\bar v(p_2)\,v(q_2)\,K_{\rm S}(k^2)\\[1ex]
&\equiv& u^\dagger(q_1)\,\gamma_0\,u(p_1)\,v^\dagger(p_2)\,\gamma_0\,v(q_2)\,
K_{\rm S}(k^2)\\[1ex]
&=&
\chi^\dagger\left(1,\frac{\mbox{\boldmath$\sigma$}\cdot{\bf q}_1}{2\,m}\right)
\left(\begin{array}{cr}1&0\\ 0&-1\end{array}\right)
\left(\begin{array}{c}1\\[1ex]
\dfrac{\mbox{\boldmath$\sigma$}\cdot{\bf p}_1}{2\,m}\end{array}\right)
\chi\\[1ex]
&\times&
{\chi^c}^\dagger\left(\frac{\mbox{\boldmath$\sigma$}\cdot{\bf p}_2}{2\,m},1
\right)
\left(\begin{array}{cr}1&0\\ 0&-1\end{array}\right)
\left(\begin{array}{c}
\dfrac{\mbox{\boldmath$\sigma$}\cdot{\bf q}_2}{2\,m}\\[2ex]
1\end{array}\right)\chi^c\,K_{\rm S}(-{\bf k}^2)\\[1ex]
&=& - \left[\chi^\dagger\,\chi - \frac{1}{4\,m^2}\,
\chi^\dagger\,(\mbox{\boldmath$\sigma$}\cdot{\bf q}_1)\,
(\mbox{\boldmath$\sigma$}\cdot{\bf p}_1)\,\chi\right]\\[1ex]
&{}&\quad\times\left[{\chi^c}^\dagger\,\chi^c - \frac{1}{4\,m^2}\,
{\chi^c}^\dagger\,(\mbox{\boldmath$\sigma$}\cdot{\bf p}_2)\,
(\mbox{\boldmath$\sigma$}\cdot{\bf q}_2)\,\chi^c\right]
K_{\rm S}(-{\bf k}^2)\\[1ex]
&=& - \left\{1 - \frac{1}{4\,m^2}\left[\chi^\dagger\,
(\mbox{\boldmath$\sigma$}\cdot{\bf q}_1)\,
(\mbox{\boldmath$\sigma$}\cdot{\bf p}_1)\,\chi
+ {\chi^c}^\dagger\,(\mbox{\boldmath$\sigma$}\cdot{\bf p}_2)\,
(\mbox{\boldmath$\sigma$}\cdot{\bf q}_2)\,\chi^c\right]\right\}\\[1ex]
&\times& K_{\rm S}(-{\bf k}^2)\ .
\end{eqnarray*}
Dropping all contributions to spin-independent relativistic corrections at
the very instant they show up, and recalling the abbreviations
(\ref{eq:sigmaexpval}), this scattering amplitude simplifies, in the
center-of-momentum frame, to
\begin{eqnarray*}
t_{\rm S}
&=& - \left[1 - \frac{1}{4\,m^2}\,(i\,\epsilon_{ijk}\,p_i\,k_j\,\sigma_{1k}
+ i\,\epsilon_{ijk}\,p_i\,k_j\,\sigma_{2k})\right]K_{\rm S}(-{\bf k}^2)\\[1ex]
&=& - \left[1 - \frac{1}{4\,m^2}\,i\,\epsilon_{ijk}\,p_i\,k_j\,
(\sigma_{1k} + \sigma_{2k})\right]K_{\rm S}(-{\bf k}^2)\ .
\end{eqnarray*}

Once more, Fourier transformation appears to be the adequate tool to obtain
the resulting effective interaction potential:
\begin{itemize}
\item For the same reason as before, the static interaction potential has to
be again spherically symmetric:
$$
V_{\rm S}(r) \equiv \frac{1}{(2\pi)^3} \int d^3k \exp(-i\,{\bf k}\cdot{\bf x})
\,K_{\rm S}(-{\bf k}^2)\ .
$$
\item The corresponding spin-dependent relativistic corrections may be found
with the help of
\begin{eqnarray*}
\frac{1}{(2\pi)^3}\int d^3k\,k_j\exp(-i\,{\bf k}\cdot{\bf x})\,
K_{\rm S}(-{\bf k}^2)
&=& i\,\nabla_j V_{\rm S}(r)\\[1ex]
&=& i\,\frac{x_j}{r}\,V'_{\rm S}(r)\ .
\end{eqnarray*}
\end{itemize}
Consequently, the spin-dependent relativistic corrections $V_{\rm spin}^{\rm
scalar}$ for the case of a scalar Lorentz structure of the effective
fermion--antifermion interaction become
$$
V_{\rm spin}^{\rm scalar} = H_{\rm LS}^{\rm scalar}\ ,
$$
with the spin--orbit term
$$
H_{\rm LS}^{\rm scalar}
= - \frac{1}{2\,m^2\,r}\,{\bf L}\cdot{\bf S}\,V_{\rm S}'(r)\ ,
$$
where ${\bf L}$ denotes, as before, the relative orbital angular momentum
of the bound-state constituents,
$$
{\bf L} \equiv {\bf x}\times{\bf p}\ .
$$
Accordingly, an interaction with scalar spin structure contributes only to
the spin--orbit term $H_{\rm LS}$. However, apart from possible differences
of the two nonrelativistic potentials $V_{\rm V}(r)$ and $V_{\rm S}(r)$ of
vector and scalar spin structure, respectively, the spin-dependent
relativistic corrections for scalar spin structure, $H_{\rm LS}^{\rm
scalar}$, contribute with a sign opposite to that of the corresponding
spin-dependent relativistic corrections for vector spin structure, $H_{\rm
LS}^{\rm vector}$. Hence, assuming identical static potentials, i.e.,
$$
V_{\rm V}(r) = V_{\rm S}(r)\ ,
$$
the spin--orbit term $H_{\rm LS}^{\rm vector}$, resulting from a vector spin
structure, may be partially compensated by the spin--orbit term $H_{\rm
LS}^{\rm scalar}$, resulting from a scalar spin structure:
$$
H_{\rm LS}^{\rm scalar} = - \frac{1}{3}\,H_{\rm LS}^{\rm vector}
\quad\mbox{for}\quad V_{\rm V}(r) = V_{\rm S}(r)\ .
$$

Before trying, in Chapter~\ref{ch:cornell}, to write down a (physically
meaningful) quark--antiquark interaction potential, it is advisable to
``condense'' all these results on the effective fermion--antifermion
interaction potential to what we would like to call a ``generalized
Breit--Fermi Hamiltonian.''

\section{Generalized Breit--Fermi Hamiltonian}\index{Breit--Fermi
Hamiltonian|(}\label{sec:breitfermi}

The Hamiltonian containing all spin-dependent relativistic corrections up to
order
$$
v^2 = \frac{{\bf p}^2}{E_p^2}\simeq\frac{{\bf p}^2}{m^2}
$$
is called the {\em generalized Breit--Fermi Hamiltonian\/}:\index{Breit--Fermi
Hamiltonian}
\begin{equation}
\fbox{\ $H = m_1 + m_2 + \dfrac{{\bf p}^2}{2\,\mu} + V_{\rm NR}(r)
+ H_{\rm LS} + H_{\rm SS} + H_{\rm T}$\ }\ ,
\label{eq:breitfermi}
\end{equation}
where
\begin{eqnarray*}
\mu &\equiv& \frac{m_1\,m_2}{m_1 + m_2}\\[1ex]
&=& \frac{m}{2}\quad\mbox{for}\quad m_1 = m_2 = m
\end{eqnarray*}
is the reduced mass and---according to the analysis of
Section~\ref{sec:spinstruc}---the static potential $V_{\rm NR}(r)$ consists
of a vector and a scalar contribution,
\begin{equation}
\fbox{\ $V_{\rm NR}(r) = V_{\rm V}(r) + V_{\rm S}(r)$\ }\ .
\label{statpot}
\end{equation}
The corresponding spin-dependent relativistic
corrections\index{spin-dependent corrections}\index{corrections!---,
spin-dependent} read (for the case of equal masses $m_1 = m_2 = m$):
\begin{itemize}
\item {\em spin--orbit term\/}:\index{spin--orbit term}
\begin{equation}
\fbox{\ $H_{\rm LS} = \dfrac{1}{2\,m^2\,r}\,{\bf L}\cdot{\bf S}
\left[3\,\frac{d}{dr}V_{\rm V}(r) - \frac{d}{dr}V_{\rm S}(r)\right]$\ }\ ,
\label{eq:hls}
\end{equation}
where
$$
{\bf S} \equiv {\bf S}_1 + {\bf S}_2
$$
is the total spin of the bound state and
$$
{\bf L} \equiv {\bf x}\times{\bf p}
$$
is the relative orbital angular momentum of its constituents;
\item {\em spin--spin term\/}:\index{spin--spin term}
\begin{equation}
\fbox{\ $H_{\rm SS} = \dfrac{2}{3\,m^2}\,{\bf S}_1\cdot{\bf S}_2\,
\Delta V_{\rm V}(r)$\ }\ ;
\label{eq:hss}
\end{equation}
\item {\em tensor term\/}:\index{tensor term}
$$
H_{\rm T} = \dfrac{1}{m^2}
\left[\dfrac{({\bf S}_1\cdot{\bf x})\,({\bf S}_2\cdot{\bf x})}{r^2}
- \dfrac{1}{3}\,{\bf S}_1\cdot{\bf S}_2\right]
\left[\dfrac{1}{r}\,\dfrac{d}{dr}V_{\rm V}(r)
- \dfrac{d^2}{dr^2}V_{\rm V}(r)\right]\ ,
$$
or, with the abbreviation
\begin{equation}
S_{12} \equiv 12\left[\dfrac{({\bf S}_1\cdot{\bf x})\,({\bf S}_2\cdot{\bf x})}
{r^2} - \dfrac{1}{3}\,{\bf S}_1\cdot{\bf S}_2\right]
\label{eq:s12}
\end{equation}
of the spin-dependent factor, which is sometimes called the tensor
operator\index{tensor operator},
\begin{equation}
\fbox{\ $H_{\rm T} = \dfrac{1}{12\,m^2}\,S_{12}
\left[\dfrac{1}{r}\,\dfrac{d}{dr}V_{\rm V}(r)
- \dfrac{d^2}{dr^2}V_{\rm V}(r)\right]$\ }\ .
\label{eq:ht}
\end{equation}
\end{itemize}
The corresponding expressions for the general case of unequal masses of the
bound-state constituents $m_1 \neq m_2$ may be obtained in a similar manner;
they are collected in Appendix~\ref{app:formspec}.

The total angular momentum ${\bf J}$ of the respective bound state under
consideration---which constitutes, of course, nothing else but the spin of
the corresponding composite particle---is clearly given by the sum of
\begin{itemize}
\item the relative orbital angular momentum
$$
{\bf L} \equiv {\bf x}\times{\bf p}
$$
and
\item the total spin
$$
{\bf S} \equiv {\bf S}_1 + {\bf S}_2
$$
\end{itemize}
of its constituents:
$$
{\bf J} \equiv {\bf L} + {\bf S}\ .
$$
Upon squaring this relation,
\begin {eqnarray*}
{\bf J}^2 &=& ({\bf L} + {\bf S})^2\\[1ex]
&=& {\bf L}^2 + {\bf S}^2 + 2\,{\bf L}\cdot{\bf S}\ ,
\end{eqnarray*}
we may express the product ${\bf L}\cdot{\bf S}$ of relative orbital angular
momentum ${\bf L}$ and total spin ${\bf S}$ in terms of the squares of ${\bf
L}$, ${\bf S}$, and ${\bf J}$,
$$
{\bf L}\cdot{\bf S}
= \dfrac{1}{2}\left({\bf J}^2 - {\bf L}^2 - {\bf S}^2\right)\ ,
$$
and, therefore, its expectation values by the corresponding expectation
values of ${\bf L}^2$, ${\bf S}^2$, and ${\bf J}^2$:
$$
\langle{\bf L}\cdot{\bf S}\rangle
= \dfrac{1}{2}\left(\langle{\bf J}^2\rangle - \langle{\bf L}^2\rangle
- \langle{\bf S}^2\rangle\right)\ .
$$
Accordingly, expressed in terms of the quantum numbers $\ell$, $S$, and $j$
of the relative orbital angular momentum ${\bf L}$, the total spin ${\bf S}$,
and the total angular momentum ${\bf J}$, respectively, denoting the bound
state, the expectation values $\langle{\bf L}\cdot{\bf S}\rangle$ of the
product of orbital angular momentum ${\bf L}$ and total spin ${\bf S}$,
originating from the spin--orbit term\index{spin--orbit term} $H_{\rm LS}$,
Eq.~(\ref{eq:hls}), read
$$
\langle{\bf L}\cdot{\bf S}\rangle = \frac{1}{2}\left[j\,(j+1) - \ell\,(\ell+1)
- S\,(S+1)\right]\ .
$$
Evidently, the expectation values of the spin--orbit term\index{spin--orbit
term} $H_{\rm LS}$, Eq.~(\ref{eq:hls}), vanish for either $\ell = 0$,
$$
\langle{\bf L}\cdot{\bf S}\rangle = 0\quad\mbox{for}\quad \ell = 0\ ,
$$
or $S = 0$,
$$
\langle{\bf L}\cdot{\bf S}\rangle = 0\quad\mbox{for}\quad S = 0\ ,
$$
contributing thus only for $\ell \neq 0$ and $S=1$. The above relation yields
the explicit nonvanishing matrix elements $\langle{\bf L}\cdot{\bf S}\rangle$
listed in Table~\ref{tab:ls}.
\begin{table}[hbt]
\caption{Nonvanishing spin--orbit couplings for $\ell \neq 0$ and
$S=1$}\label{tab:ls}\index{spin--orbit coupling}
$$
\begin{array}{ccc}\hline\hline\\[-1.5ex]
j&\quad&\langle{\bf L}\cdot{\bf S}\rangle\\[1ex] \hline\\[-1.5ex]
\ell+1&\quad&\ell\\[1ex]
\ell&\quad&-1\\[1ex]
\ell-1&\quad&-(\ell +1)\\[1ex] \hline\hline
\end{array}
$$
\end{table}

For fermionic bound-state constituents of spin
$$
S_1 = S_2 = \frac{1}{2}\ ,
$$
the expectation values $\langle{\bf S}_1\cdot{\bf S}_2\rangle$ of the product
of their spins ${\bf S}_1$ and ${\bf S}_2$ in the spin--spin
term\index{spin--spin term} $H_ {\rm SS}$, Eq.~(\ref{eq:hss}), have been
determined already in Section~\ref{sec:spinstruc}:
$$
\langle{\bf S}_1\cdot{\bf S}_2\rangle = \left \{\begin{array}{rl}
-\dfrac{3}{4}&\quad\mbox{for spin singlets, i.e.,}\ S = 0\ ,\\[2ex]
+\dfrac{1}{4}&\quad\mbox{for spin triplets, i.e.,}\ S = 1\ .\end{array}\right.
$$

Likewise, for fermionic bound-state constituents of spin
$$
S_1 = S_2 = \frac{1}{2}\ ,
$$
the spin-dependent factor $S_{12}$ in the tensor term\index{tensor term}
$H_{\rm T}$, Eq.~(\ref{eq:ht}), may be rewritten as
\begin{equation}
S_{12} = 2\left[3\,\dfrac{({\bf S}\cdot{\bf x})^2}{r^2} - {\bf S}^2\right]\ .
\label{eq:s12new}
\end{equation}
Accordingly, the expectation values of the tensor term $H_{\rm T}$,
Eq.~(\ref{eq:ht}), also vanish for either $S=0$,
$$
\langle S_{12}\rangle = 0\quad\mbox{for}\quad S = 0\ ,
$$
or $\ell = 0$,
$$
\langle S_{12}\rangle = 0\quad\mbox{for}\quad \ell = 0\ ,
$$
the (more or less obvious) reason for the latter being the fact that the
rotational symmetry realized in the case of $\ell = 0$ leads to
$$
\left\langle\dfrac{x_i\,x_j}{r^2}\right\rangle = \dfrac{1}{3}\,\delta_{ij}\ ,
$$
which, in turn, results in a mutual cancellation of the two terms on the
right-hand side of Eq.~(\ref{eq:s12new}). After a lengthy calculation, the
following expression for the {\em diagonal\/} matrix elements of $S_{12}$ may
be found:
$$
\langle S_{12}\rangle = \frac{4}{(2\,\ell+3)\,(2\,\ell-1)}\left[
\langle{\bf S}^2\rangle\,\langle{\bf L}^2\rangle
- \frac{3}{2}\,\langle{\bf L}\cdot{\bf S}\rangle
- 3\left(\langle{\bf L}\cdot{\bf S}\rangle\right)^2\right]\ ,
$$
which, again only for $\ell \ne 0$ and $S=1$, yields the explicit
nonvanishing expectation values listed in Table~\ref{tab:s12}.
\begin{table}[hbt]
\caption{Nonvanishing diagonal matrix elements of $S_{12}$ for $\ell \neq 0$
and $S=1$}\label{tab:s12}
$$
\begin{array}{ccc}\hline\hline\\[-1.5ex]
j&\quad&\langle S_{12}\rangle\\[1ex] \hline\\[-1.5ex]
\ell+1&\quad&-\dfrac{2\,\ell}{2\,\ell+3}\\[2ex]
\ell&\quad&2\\[1ex]
\ell-1&\quad&-\dfrac{2\,(\ell + 1)}{2\,\ell -1}\\[2ex] \hline\hline
\end{array}
$$
\end{table}
\index{relativistic corrections|)}\index{corrections!---,
relativistic|)}\index{spin-dependent corrections|)}\index{corrections!---,
spin-dependent|)}\index{Breit--Fermi Hamiltonian|)}

\chapter{The Prototype}\label{ch:cornell}

We are now in a position where we may start to think seriously about the
question of how a realistic, that is, phenomenologically acceptable,
potential describing the forces acting between quarks might look like.

\section{Funnel potential}\index{funnel potential|(}

To begin with, let's summarize our knowledge gained so far. According to the
analysis of Section~\ref{sec:spinstruc}, the quark--antiquark potential
$V_{\rm NR}(r)$ is of vector and/or scalar type,
$$
V_{\rm NR}(r) = V_{\rm V}(r) + V_{\rm S}(r)\ .
$$
For short distances, the potential---arising from one-gluon exchange---is
(essentially) Coulomb-like,
$$
V_{\rm exch}(r) = - \frac{4}{3}\,\frac{\alpha_{\rm s}}{r} \ .
$$
For large distances, there has to exist a contribution $V_{\rm conf}(r)$ in
order to describe colour confinement,
$$
V_{\rm conf}(r) = a\,r^n\quad\mbox{with}\quad n > 0\ ,
$$
implying that for large $r$ the binding force $K$ must not decrease faster
than $1/r$:
$$
K = - \frac{d}{dr}V_{\rm conf}(r) = -n\,a\,r^{n-1}
= - n\,\frac{a}{r^{1-n}}\ .
$$
From the mesonic mass spectrum, the exponent $n$ is in the vicinity of
$$
n \simeq 1\ .
$$
For instance, $n=2$ corresponds to the harmonic oscillator and would lead to
equidistant level spacings. Moreover, lattice gauge theories also find that
$V_{\rm NR}(r)$ is roughly proportional to $r$ for large $r$. Consequently, a
linear potential,
$$
V_{\rm conf}(r) = a\,r\ ,
$$
is beyond doubt a sensible choice for $V_{\rm conf}(r)$.

\begin{figure}[htb]
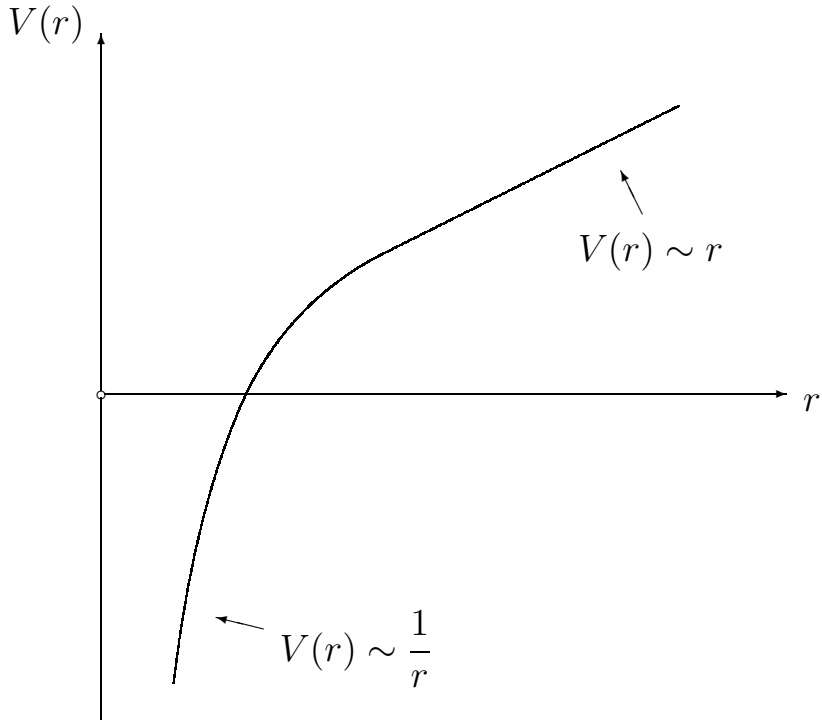

\begin{center}
\trichter
\end{center}
\caption{Funnel potential.}\label{fig:funnel}
\end{figure}

The funnel (or ``Coulomb--plus--linear'') potential (Fig.~\ref{fig:funnel})
\begin{equation}
V_{\rm NR}(r)
= \underbrace{-\frac{4}{3}\,\frac{\alpha_{\rm s}}{r}}
_{\mbox{one-gluon exchange}}
+ \underbrace{a\,r}_{\mbox{confinement}}
\label{eq:cornell-pot}
\end{equation}
fixed in this way has been the first proposed model \cite{eichten75}, which
in spite of its simplicity is able to reproduce quite well the charmonium
spectrum. In a strict sense, the momentum (-transfer) dependence
$$
\alpha_{\rm s} = \alpha_{\rm s}(Q^2)
$$
of the strong fine structure constant $\alpha_{\rm s}$ has to be taken into
account, modifying thereby the Coulomb-like behaviour of the first term on
the right-hand side of Eq.~(\ref{eq:cornell-pot}).

\section{Lorentz structure of the funnel potential}\index{funnel
potential!---, spin structure|(}\index{funnel potential!---, Lorentz
structure|(}

In order to decide whether the Lorentz structure of the funnel potential
(\ref{eq:cornell-pot}) is a pure vector, or a pure scalar, or a mixing of
both, we consider the P-wave spin splittings of charmonium and bottomonium,
that is, the ratio of mass differences \cite{schnitzer75}
\begin{equation}
\rho = \frac
{M({}^3\mbox{P}_2) - M({}^3\mbox{P}_1)}
{M({}^3\mbox{P}_1) - M({}^3\mbox{P}_0)}\ .
\label{eq:rho}
\end{equation}
(Recall the usual spectroscopic notation, which designates a state with
orbital angular momentum $\ell$, spin $S$, and total angular momentum $j$ by
${}^{2S+1}\mbox{L}_j$, where the capital $\mbox{L} \equiv
\mbox{S},\mbox{P},\mbox{D},\mbox{F},\dots$ represents the orbital angular
momentum $\ell=0,1,2,3,\dots\;$, respectively.)
\begin{table}[htb]
\caption[]{Masses (in GeV) and the ratio $\rho$ [Eq.~(\ref{eq:rho})] for the
($\ell = 1, S = 1$) states of charmonium and bottomonium
\cite{partprop94}}\label{tab:rho}
\begin{center}
\begin{tabular}{llll}\hline\hline\\[-1.5ex]
$\mbox{Level}$&$\quad ({\rm c}\bar{\rm c})$&$\quad ({\rm b}\bar{\rm b})$&
$\quad ({\rm b}\bar{\rm b})'$
\\[1ex]\hline\\[-1.5ex]
${}^3\mbox{P}_0$&$\quad 3.4151\quad$&$\quad 9.8598\quad$&$\quad 10.2321$\\[1ex]
${}^3\mbox{P}_1$&$\quad 3.5105\quad$&$\quad 9.8919\quad$&$\quad 10.2552$\\[1ex]
${}^3\mbox{P}_2$&$\quad 3.5562\quad$&$\quad 9.9132\quad$&$\quad 10.2685$\\[1ex]
\hline\\[-1.5ex]
$\rho$&$\quad 0.478\quad$&$\quad 0.664\quad$&$\quad 0.576$\\[1ex]
\hline\hline
\end{tabular}
\end{center}
\end{table}

From Table~\ref{tab:rho}, the experimental average for $\rho$
\cite{partprop94},
$$
\begin{array}{l}
\rho = 0.48\quad\mbox{for}\quad ({\rm c}\bar{\rm c})\ ,\\[1ex]
\rho = 0.66\quad\mbox{for}\quad ({\rm b}\bar{\rm b})\ ,\\[1ex]
\rho = 0.58\quad\mbox{for}\quad ({\rm b}\bar{\rm b})'\ ,
\end{array}
$$
is
$$
\rho_{\rm exp} \simeq 0.6\ .
$$

With the help of the generalized Breit--Fermi Hamiltonian
(\ref{eq:breitfermi}), we calculate this ratio $\rho$ for the potential
(\ref{eq:cornell-pot}) perturbatively. Since the spin--spin interaction,
$H_{\rm SS}$, does not depend on the total angular momentum, its contribution
cancels in a perturbative evaluation of $\rho$. Accordingly, $\rho$ is
determined by the contributions of spin--orbit term, $H_{\rm LS}$, and tensor
term, $H_{\rm T}$, only:
$$
\rho = \frac{
\langle{}^3\mbox{P}_2|H_{\rm LS} + H_{\rm T}|{}^3\mbox{P}_2\rangle -
\langle{}^3\mbox{P}_1|H_{\rm LS} + H_{\rm T}|{}^3\mbox{P}_1\rangle}{
\langle{}^3\mbox{P}_1|H_{\rm LS} + H_{\rm T}|{}^3\mbox{P}_1\rangle -
\langle{}^3\mbox{P}_0|H_{\rm LS} + H_{\rm T}|{}^3\mbox{P}_0\rangle}\ .
$$
From Tables~\ref{tab:ls} and \ref{tab:s12}, we find for the expectation
values $\langle{\bf L}\cdot{\bf S}\rangle$ and $\langle S_{12}\rangle$:
$$
\langle{\bf L}\cdot{\bf S}\rangle = \left\{\begin{array}{rc}
-2&\quad\mbox{for}\quad{}^3\mbox{P}_0\ ,\\[1ex]
-1&\quad\mbox{for}\quad{}^3\mbox{P}_1\ ,\\[1ex]
 1&\quad\mbox{for}\quad{}^3\mbox{P}_2\ ,
\end{array}\right.
$$
and
$$
\langle S_{12}\rangle = \left\{\begin{array}{rc}
-4&\quad\mbox{for}\quad{}^3\mbox{P}_0\ ,\\[1ex]
2&\quad\mbox{for}\quad{}^3\mbox{P}_1\ ,\\[1ex]
-\dfrac{2}{5}&\quad\mbox{for}\quad{}^3\mbox{P}_2\ .
\end{array}\right.
$$
\begin{itemize}
\item For a pure vector, i.e.,
$$V_{\rm V} = V_{\rm NR}\ ,\quad V_{\rm S}=0\ ,
$$
one obtains
$$
\rho = \frac{1}{5}\,\frac{
8\,\alpha_{\rm s}\,\langle r^{-3}\rangle + 7\,a\,\langle r^{-1}\rangle}{
2\,\alpha_{\rm s}\,\langle r^{-3}\rangle + a\,\langle r^{-1}\rangle}\ ,
$$
which implies the bounds
$$
\frac{4}{5}\leq\rho\leq\frac{7}{5}
$$
corresponding to $a = 0$ and $\alpha_{\rm s}=0$, respectively, in clear
conflict with the experimental finding $\rho_{\rm exp}\simeq 0.6$.
\item A pure scalar, i.e.,
$$
V_{\rm S} = V_{\rm NR}\ ,\quad V_{\rm V}=0\ ,
$$
leads to
$$
\rho = 2\ ,
$$
which also is not tolerable from an experimental point of view.
\item A vector/scalar mixing, i.e.,
$$
V_{\rm V} = V_{\rm exch} = -\dfrac{4}{3}\,\dfrac{\alpha_{\rm s}}{r}\ ,\quad
V_{\rm S} = V_{\rm conf} = a\,r\ ,
$$
results in
$$
\rho = \frac{1}{5}\,
\frac{8\,\alpha_{\rm s}\,\langle r^{-3}\rangle -
\dfrac{5}{2}\,a\,\langle r^{-1}\rangle}
{2\,\alpha_{\rm s}\,\langle r^{-3}\rangle -
\dfrac{1}{4}\,a\,\langle r^{-1}\rangle}\ ,
$$
which implies
$$
\rho\leq\frac{4}{5}\ ,
$$
if the Coulomb part dominates, and
$$
\rho\geq 2\ ,
$$
if the linear part dominates.
\end{itemize}

Consequently, we arrive at the conclusion that the funnel potential $V_{\rm
NR}$ in Eq.~(\ref{eq:cornell-pot}) must be a linear combination of a vector
and a scalar part,
$$
V_{\rm NR}(r) = V_{\rm V}(r) + V_{\rm S}(r)\ ,
$$
where the Coulomb part $V_{\rm exch}$ is of vector type,
$$
V_{\rm V}(r) = V_{\rm exch}(r) = - \dfrac{4}{3}\,\dfrac{\alpha_{\rm s}}{r}\ ,
$$
and the linear part $V_{\rm conf}$ is of scalar type,
$$
V_{\rm S}(r) = V_{\rm conf}(r) = a\,r\ .
$$

In summary, from the analysis of the most general conceivable spin structures
and the experimentally observed quarkonium mass spectra, we have been able to
determine unambiguously the basic shape of the potential acting between
quarks:
\begin{center}
\begin{tabular}{|c|}\hline\\[-2ex]
{\bf The interquark potential $V_{\rm NR}(r) = V_{\rm V}(r) + V_{\rm S}(r)$
essentially}\\
{\bf consists of a Coulomb part $V_{\rm exch}$, which is of vector
type,}\\[1ex]
{\bf $V_{\rm V}(r) = V_{\rm exch}(r) = - \dfrac{4}{3}\,\dfrac{\alpha_{\rm
s}}{r}\ $,}\\[2ex]
{\bf as well as of a linear part $V_{\rm conf}$, which is of scalar
type,}\\[1ex]
{\bf $V_{\rm S}(r) = V_{\rm conf}(r) = a\,r\ $.}\\[1ex]\hline
\end{tabular}
\end{center}

In this form, the funnel potential represents the genuine prototype of all
``QCD-inspired'' potential models proposed for the description of hadrons as
bound states of (``constituent'') quarks \cite{lucha89bi,lucha91,lucha92}. A
selection of more sophisticated potential models may be found in
Appendix~\ref{ch:varpot}.\index{funnel potential|)}\index{funnel
potential!---, spin structure|)}\index{funnel
potential!---, Lorentz structure|)}

\appendix

\chapter{S Matrix, Cross-section, and Decay Width}\index{S
matrix|(}\index{S-matrix element|(}\index{cross-section|(}\index{decay
width|(}\index{decay rate|(}\label{app:smatrix}

The normalization of creation and annihilation operators is reflected by the
(anti-) commutation relations of these operators:
\begin{itemize}
\item For the case of bosons, the nonvanishing commutators are
$$
[a({\bf p}),a^\dagger({\bf q})] = \delta^{(3)}({\bf p} - {\bf q})\ .
$$
\item For the case of fermions, the nonvanishing anticommutators are
$$
\{b({\bf p},\sigma),b^\dagger({\bf q},\tau)\} =
\{d({\bf p},\sigma),d^\dagger({\bf q},\tau)\} =
\delta^{(3)}({\bf p} - {\bf q})\,\delta_{\sigma\tau}\ .
$$
\end{itemize}
Normalizing the vacuum state $|0\rangle$ according to
$$
\langle 0|0\rangle = 1\ ,
$$
the normalizations of the one-particle states\index{one-particle state!---,
normalization} read
\begin{itemize}
\item for bosons, generically denoted by ${\rm B}$,
$$
\langle{\rm B}({\bf p})|{\rm B}({\bf q})\rangle
= \delta^{(3)}({\bf p} - {\bf q})
$$
and
\item for fermions, generically denoted by ${\rm F}$,
$$
\langle{\rm F}({\bf p},\sigma)|{\rm F}({\bf q},\tau)\rangle
= \delta^{(3)}({\bf p} - {\bf q})\,\delta_{\sigma\tau}\ .
$$
\end{itemize}

Let us define, for any transition ${\rm i}\rightarrow{\rm f}$ from some
initial state ${\rm i}$ to some final state ${\rm f}$, like, for instance, a
scattering or decay process, the {\em S-matrix element\/}\index{S-matrix
element} $S_{\rm f\/i}$ by
$$
S_{\rm f\/i} \equiv \langle{\rm f},\mbox{out}|{\rm i},\mbox{in}\rangle =
\langle{\rm f}|S|{\rm i}\rangle
$$
and the {\em reduced T-matrix element\/}\index{T-matrix element} $T_{\rm
f\/i}$ by
$$
S_{\rm f\/i} =: \delta_{\rm f\/i}
+ i\,(2\pi)^4\,\delta^{(4)}(P_{\rm f} - P_{\rm i})\,T_{\rm f\/i}\ ,
$$
where $P_{\rm i}$ and $P_{\rm f}$ denote the total momenta of initial and
final state, respectively. The corresponding {\em transition
probability\/}\index{transition probability} $W_{\rm f\/i}$ is the square of
the modulus of the transition amplitude $S_{\rm f\/i} - \delta_{\rm f\/i}$:
$$
W_{\rm f\/i} \equiv |S_{\rm f\/i} - \delta_{\rm f\/i}|^2\ .
$$
For a finite spatial volume $V$ and a finite time interval $T$,
the obscure square of the $\delta$ function may be replaced by
\begin{eqnarray*}
\left[(2\pi)^4\,\delta^{(4)}(P_{\rm f} - P_{\rm i})\right]^2
&=& (2\pi)^4\,\delta^{(4)}(P_{\rm f} - P_{\rm i})\int d^4x
\exp[i\,(P_{\rm f} - P_{\rm i})\,x]\\[1ex]
&=& (2\pi)^4\,\delta^{(4)}(P_{\rm f} - P_{\rm i})\,V\,T\ ,
\end{eqnarray*}
which leads to
$$
W_{\rm f\/i}
= (2\pi)^4\,\delta^{(4)}(P_{\rm f} - P_{\rm i})\,V\,T\,|T_{\rm f\/i}|^2\ .
$$
The {\em transition rate\/}\index{transition rate} $R_{\rm f\/i}$ is the
transition probability per unit time:
$$
\fbox{\ $R_{\rm f\/i} \equiv \dfrac{W_{\rm f\/i}}{T}
= (2\pi)^4\,\delta^{(4)}(P_{\rm f} - P_{\rm i})\,V\,|T_{\rm f\/i}|^2$\ }\ .
$$

The {\em cross-section\/}\index{cross-section} $\sigma_{\rm f\/i}$ is the
above transition rate $R_{\rm f\/i}$ divided by the product of the observed
flux of the incoming particles, $j = n\,v_{\rm rel}$, times the number of
target particles, $n\,V$; here, $v_{\rm rel}$ is the relative velocity of the
scattered particles and $n$ denotes generically the particle densities. If
necessary, one has to sum over the final states and to average over the
initial states, which will be indicated below by a primed sum over the
possible spin polarizations $\sigma$:
$$
\sigma_{\rm f\/i} \equiv \dfrac{1}{v_{\rm rel}\left(\prod_{i=1,2}n_i\right)V}
\int \left(\prod_f d^3p_f\right){\sum_\sigma}'R_{\rm f\/i}\ .
$$
With the particle density
$$
n = \frac{1}{(2\pi)^3}\ ,
$$
corresponding to a normalization volume of size $(2\pi)^3$, the resulting
cross-section\index{cross-section} $\sigma({\rm i}\rightarrow{\rm f})$ for the
scattering process ${\rm i}\rightarrow{\rm f}$ reads
$$
\fbox{\ $\displaystyle\sigma({\rm i}\rightarrow{\rm f})
= \dfrac{(2\pi)^{10}}{v_{\rm rel}} \int\left(\prod_f d^3p_f\right)
\delta^{(4)}(P_{\rm f} - P_{\rm i})\,{\sum_\sigma}'|T_{\rm f\/i}|^2$\ }\ .
$$
The product of the energies $E_1$ and $E_2$ of the two scattered particles
and of their relative velocity $v_{\rm rel}$ forms a Lorentz invariant:
$$
E_1\,E_2\,v_{\rm rel} = \sqrt{(p_1\,p_2)^2 - m_1^2\,m_2^2}\ .
$$

The {\em decay width\/}\index{decay width}\index{decay rate} $\Gamma_{\rm
f\/i}$ is the transition rate $R_{\rm f\/i}$ divided by the number of
decaying particles $n\,V$. If necessary, one again has to sum over the final
states and to average over the initial states:
$$
\Gamma_{\rm f\/i} \equiv \frac{1}{n_{\rm i}\,V}\int\left(\prod_f d^3p_f\right)
{\sum_\sigma}'R_{\rm f\/i}\ .
$$
The {\em partial decay width\/}\index{decay width!---, partial}\index{partial
decay width}\index{decay rate!---, partial}\index{partial decay rate}
$\Gamma({\rm i}\rightarrow{\rm f})$ for the decay of the particle ${\rm i}$
into a particular final state ${\rm f}$ is therefore given by
$$
\fbox{\ $\displaystyle\Gamma({\rm i}\rightarrow{\rm f})
= (2\pi)^7\int\left(\prod_f d^3p_f\right)\delta^{(4)}(P_{\rm f} - P_{\rm i})\,
{\sum_\sigma}'|T_{\rm f\/i}|^2$\ }\ .
$$
The {\em total decay width\/}\index{decay width!---, total}\index{total decay
width}\index{decay rate!---, total}\index{total decay rate} $\Gamma({\rm i})$
of the particle ${\rm i}$, which is, of course, nothing else but the inverse
of the average lifetime\index{lifetime} $\tau_{\rm i}$ of this particle, is
obtained by summing over all possible, kinematically allowed decay channels
${\rm f}$:
$$
\Gamma({\rm i}) \equiv \frac{1}{\tau_{\rm i}}
= \sum_{\rm f}\Gamma({\rm i}\rightarrow{\rm f})\ .
$$
\index{S matrix|)}\index{S-matrix
element|)}\index{cross-section|)}\index{decay width|)}\index{decay rate|)}

\chapter{Feynman Rules for a General Gauge Theory}\index{Feynman
rules|(}\index{gauge theory|(}\label{app:feynman}

First of all, a little {\em warning\/}: The correct application of Feynman
rules requires some experience. In particular, one clearly should be careful
\begin{itemize}
\item when identifying all Feynman diagrams regarded as relevant for the
specific process under consideration and
\item when computing the combinatorial factors (cf. rule \# 3 below).
\end{itemize}
In order to remain on the safe side, it is advisable to evaluate $n$-point
Green's functions with the help of ``Wick's theorem.'' Wick's theorem allows
to convert time-ordered products of field operators, like those appearing in
the S operator\index{S operator}
$$
S = T \exp\left[i\int d^4x\,{\cal L}_{\rm I}(x)\right]\ ,
$$
into a sum of products of propagators and normal-ordered products of field
operators. A particular Feynman graph is then nothing else but the symbolical
representation of a particular operator in the series of the Wick
decomposition. S-matrix elements may be obtained from the Green's functions
along the course of the ``LSZ reduction technique.''

A general unbroken non-Abelian gauge theory for Dirac fermions $\psi$ but
without scalar bosons is described in R${}_\xi$ gauge by the
Lagrangian\index{Lagrangian!---, general gauge-invariant}
$$
{\cal L} = - \frac{1}{4}\,F^a_{\mu\nu}\,F^{\mu\nu}_a
+ \bar\psi\left(i\,\Dsla - m\right)\psi
- \frac{\xi}{2}\left(\partial_\mu V^\mu_a\right)^2
+ \left(\partial_\mu\bar\zeta\right) D^\mu\zeta \ ,
$$
$$
F^a_{\mu\nu} \equiv \partial_\mu V^a_\nu - \partial_\nu V^a_\mu
+ g\,f_{abc}\,V^b_\mu\,V^c_\nu \ ,
$$
$$
D_\mu \equiv \partial_\mu - i\,g\,V^a_\mu\,T^a \ .
$$
The fermions $\psi$ transform according to an arbitrary representation of the
gauge group.

\newpage

The modification of the original theory brought about by the gauge
fixing\index{gauge fixing} must be compensated---in order to maintain the
unitarity of the S matrix\index{S matrix}\index{S matrix!---, unitarity}---by
adding a further (in general, not Hermitean) term to the Lagrangian, which
involves anticommuting, scalar ``ghost''\index{ghost} fields $\zeta_a$,
$\bar\zeta_a$.

For this theory, the complete list of Feynman rules in momentum space for the
computation of $i\,T_{\rm f\/i}$, where $T_{\rm f\/i}$ is the reduced
T-matrix element defined (in Appendix~\ref{app:smatrix}) in terms of the
S-matrix element $S_{\rm f\/i}$ by
$$
S_{\rm f\/i} \equiv \langle{\rm f},\mbox{out}|{\rm i},\mbox{in}\rangle
=: \delta_{\rm f\/i}
+ i\,(2\pi)^4\,\delta^{(4)}(P_{\rm f} - P_{\rm i})\,T_{\rm f\/i}\ ,
$$
is:
\begin{table}[htb]
1. {\em Propagators\/}:
\caption{Feynman rules for the propagators in a general gauge
theory}\index{Feynman rules}\index{propagator}\index{gauge
theory}\index{vector-boson propagator}\index{fermion propagator}\index{ghost
propagator}
$$
\begin{array}{|c|l|}\hline
&\\[-2ex]
\quad\mbox{Vector-boson propagator}\quad&\quad
{\cal L}_0
= -\dfrac{1}{4}\left(\partial_\mu V^a_\nu - \partial_\nu V^a_\mu\right)^2
- \dfrac{\xi}{2}\left(\partial_\mu V^\mu_a\right)^2\quad\\[2ex]
\hline
&\\[-1ex]
&\quad = i\,D_{\rm F}(k)^{ab}_{\mu\nu}\ ,\\[1ex]
\vecprop&\quad D_{\rm F}(k)^{ab}_{\mu\nu} = \dfrac{1}{k^2 + i\,\epsilon}
\left[-g_{\mu\nu}\dfrac{}{}\right.\quad\\[2ex]
&\quad\hspace{72pt} + \left.\left(1-\dfrac{1}{\xi}\right)
\dfrac{k_\mu\,k_\nu}{k^2 + i\,\epsilon}\right]\delta_{ab}\quad\\[3ex]
\hline\hline
&\\[-1ex]
\mbox{Fermion propagator}&\quad
{\cal L}_0 = \bar\psi\left(i\,\dsla - m\right)\psi\\[2ex]
\hline
&\\[-1ex]
\ferprop&\quad = i\,S_{\rm F}(k)\ ,\\[1ex]
&\quad S_{\rm F}(k) = \dfrac{1}{\ksla - m + i\,\epsilon} \equiv
\dfrac{\ksla + m}{k^2 - m^2 + i\,\epsilon}\quad\\[3ex]
\hline\hline
&\\[-1ex]
\mbox{Ghost propagator}&\quad
{\cal L}_0 = \left(\partial_\mu\bar\zeta_a\right)\partial^\mu\zeta_a\\[2ex]
\hline
&\\[-1ex]
\ghprop&\quad = i\,\Delta_{\rm F}(k)_{ab}\ ,\\[1ex]
&\quad \Delta_{\rm F}(k)_{ab}
= \dfrac{1}{k^2 + i\,\epsilon}\,\delta_{ab}\\[3ex]
\hline
\end{array}
$$
\end{table}
\clearpage
\begin{table}[htb]
2. {\em Vertices\/}:
\caption{Feynman rules for the vertices in a general gauge
theory}\index{Feynman rules}\index{vertex}\index{gauge
theory}\index{three--vector-boson vertex}\index{four--vector-boson
vertex}\index{vector-boson--fermion vertex}\index{vector-boson--ghost vertex}
$$
\begin{array}{|c|l|}\hline
&\\[-1ex]
\mbox{Three--vector-boson vertex}&\quad
{\cal L}_{\rm I} = -g\,f_{abc}\,V^\mu_a\,V^\nu_b\,\partial_\mu V^c_\nu\\[2ex]
\hline
&\\[-1ex]
&\\
&\quad = - g\,f_{abc}\left[g_{\mu\nu}\,(p - q)_\rho\right.\quad\\[1ex]
\threevert
&\quad\hspace{56pt}\ + \left.g_{\nu\rho}\,(q - r)_\mu\right.\quad\\[1ex]
&\quad\hspace{56pt}\ + \left.g_{\rho\mu}\,(r - p)_\nu\right]\quad\\
&\\[-1ex]
&\\
\hline\hline
&\\[-2ex]
\mbox{Four--vector-boson vertex}&\quad
{\cal L}_{\rm I}
= -\dfrac{1}{4}\,g^2\,f_{abc}\,f_{ade}\,
V^b_\mu\,V^c_\nu\,V^\mu_d\,V^\nu_e\\[2ex]
\hline
&\\[-1ex]
&\\
&\quad = - i\,g^2\left[f_{abe}\,f_{cde}\,
(g_{\mu\rho}\,g_{\nu\sigma} - g_{\mu\sigma}\,g_{\nu\rho})\right.\quad\\[1ex]
\fourvert
&\quad\hspace{45pt}\ + \left.f_{ace}\,f_{bde}\,
(g_{\mu\nu}\,g_{\rho\sigma} - g_{\mu\sigma}\,g_{\rho\nu})\right.\quad\\[1ex]
&\quad\hspace{45pt}\ + \left.f_{ade}\,f_{bce}\,
(g_{\mu\nu}\,g_{\sigma\rho} - g_{\mu\rho}\,g_{\sigma\nu})\right]\quad\\
&\\[-1ex]
&\\
\hline\hline
&\\[-1ex]
\quad \mbox{Vector-boson--fermion vertex} \quad&\quad
{\cal L}_{\rm I} = g\,\bar\psi\,\Vsla\,\psi\\[2ex]
\hline
&\\[-1ex]
&\\
&\\
\fervert&\quad = i\,g\,\gamma_\mu\,T^a_{ij}\\
&\\
&\\[-1ex]
&\\
\hline\hline
&\\[-1ex]
\mbox{Vector-boson--ghost vertex}&\quad
{\cal L}_{\rm I}
= -g\,f_{abc}\left(\partial_\mu\bar\zeta_a\right)\zeta_b\,V^\mu_c\\[2ex]
\hline
&\\[-1ex]
&\\
&\\
\ghvert&\quad = g\,f_{abc}\,p_\mu\\
&\\
&\\[-1ex]
&\\
\hline
\end{array}
$$
\end{table}
\clearpage
\noindent
3. {\em Symmetry factors\/}:\index{symmetry factor}\index{combinatorial
factor}\index{statistical factor} For Feynman graphs involving identical
particles in internal lines there arise certain combinatorial factors, which
have to be introduced in order to avoid ``double counting.'' Some examples
for this are presented in Table~\ref{tab:combfac}.
\begin{table}[htb]
\caption{Combinatorial factors for Feynman diagrams involving identical
particles in internal lines}\index{symmetry factor}\index{combinatorial
factor}\index{statistical factor}\label{tab:combfac}
$$
\begin{array}{|c|c|}\hline
&\\[-1ex]
\qquad\qquad\mbox{Feynman graph}\qquad\qquad&\quad
\mbox{Statistical factor}\quad\\[2ex]
\hline
&\\[-2ex]
&\\
&\\
\combtad&\dfrac{1}{2!}\\
&\\
&\\[-2ex]
&\\
\hline
&\\[-2ex]
&\\
&\\
\combone&\dfrac{1}{2!}\\
&\\
&\\[-2ex]
&\\
\hline
&\\[-2ex]
&\\
&\\
\combtwo&\dfrac{1}{3!}\\
&\\
&\\[-2ex]
&\\
\hline
\end{array}
$$
\end{table}

\noindent
4. For each closed loop\index{closed loop}\index{loop!---, closed} of
anticommuting fields, i.e., fermions or ghosts, a factor $$(-1)\ .$$
5. {\em Loop integration\/}:\index{loop integration} At every vertex,
energy-momentum conservation has to be taken into account. For every internal
and independent four-momentum $k$, i.e., one which is {\em not\/} constrained
by energy-momentum conservation at the vertices, an integration
$$
\int\frac{d^4k}{(2\pi)^4} \ .
$$

\newpage

\noindent
6. {\em External particles\/}:\index{external particle}\index{particle!---,
external} Let the polarization vectors\index{polarization vector!---,
normalization} $\epsilon_\mu(p,\lambda)$ describing (massless) vector bosons
$V_\mu$ of four-momentum $p$ and spin polarization $\lambda$ be normalized
according to
$$
\sum_\lambda\epsilon_\mu(p,\lambda)\,\epsilon^\ast_\nu(p,\lambda)
= - g_{\mu\nu}\ .
$$
Let the Dirac spinors\index{Dirac spinor!---, normalization} $u(p,\sigma)$
and $v(p,\sigma)$ describing fermions $\psi$ of mass $m$, four-momentum $p$,
and spin polarization $\sigma$ be normalized according to
$$
\begin{array}{rcl}
\bar u(p,\sigma)\,u(p,\tau) &=&\msp \delta_{\sigma\tau} \ ,\\[1ex]
\bar v(p,\sigma)\,v(p,\tau) &=& - \delta_{\sigma\tau} \ ,
\end{array}
$$
which is equivalent to
$$
u^\dagger(p,\sigma)\,u(p,\tau) = v^\dagger(p,\sigma)\,v(p,\tau)
= \frac{E_p}{m}\,\delta_{\sigma\tau} \ ,
$$
where
$$
E_p \equiv \sqrt{{\bf p}^2 + m^2}\ .
$$
The above normalization implies for the energy projection
operators\index{energy projection operator}
$$
\begin{array}{rcl}
{\displaystyle \sum_\sigma}u(p,\sigma)\,\bar u(p,\sigma) &=& \dfrac{\psla +
m}{2\,m} \ ,\\[2ex]
{\displaystyle \sum_\sigma}v(p,\sigma)\,\bar v(p,\sigma) &=& \dfrac{\psla -
m}{2\,m} \ .
\end{array}
$$
With the above conventions, the expansions of the corresponding field
operators\index{field operator!---, plane-wave expansion} in terms of plane
waves read
\begin{itemize}
\item for Hermitean vector bosons
$$
\begin{array}{r}
V_\mu(x) = \dfrac{1}{(2\pi)^{3/2}}\int\frac{d^3p}{\sqrt{2\,E_p}}\,\sum_\lambda
\left[a(p,\lambda)\,\epsilon_\mu(p,\lambda)\exp(-i\,p\,x)\ \right.\\
+ \left.a^\dagger(p,\lambda)\,\epsilon^\ast_\mu(p,\lambda)\exp(i\,p\,x)\right]
\end{array}
$$
and
\item for Dirac fermions
$$
\begin{array}{r}
\psi(x)
= \dfrac{1}{(2\pi)^{3/2}}\int d^3p\,\sqrt{\dfrac{m}{E_p}}\,\sum_\sigma
\left[b(p,\sigma)\,u(p,\sigma)\exp(-i\,p\,x)\quad\right.\\
+ \left.d^\dagger(p,\sigma)\,v(p,\sigma)\exp(i\,p\,x)\right] \ ,\\[3ex]
\bar\psi(x)
= \dfrac{1}{(2\pi)^{3/2}}\int d^3p\,\sqrt{\dfrac{m}{E_p}}\,\sum_\sigma
\left[d(p,\sigma)\,\bar v(p,\sigma)\exp(-i\,p\,x)\quad\right.\\
+ \left.b^\dagger(p,\sigma)\,\bar u(p,\sigma)\exp(i\,p\,x)\right] \ .
\end{array}
$$
\end{itemize}

\newpage

\begin{table}[htb]
\caption{Feynman rules for the external particles in a general gauge
theory}\index{Feynman rules}\index{external particle}\index{particle!---,
external}
\begin{center}
\begin{tabular}{|l|c|c|}\hline
&&\\
\multicolumn{1}{|c|}{Particle}& incoming&outgoing \\
&&\\
\hline
&&\\
$\quad$ Vector boson$\quad$ &$\quad
\dfrac{1}{(2\pi)^{3/2}\,\sqrt{2\,E_p}}\,\epsilon_\mu(p,\lambda)\quad$&$\quad
\dfrac{1}{(2\pi)^{3/2}\,\sqrt{2\,E_p}}\,\epsilon^\ast_\mu(p,\lambda)\quad$ \\
&&\\
\hline
&&\\
$\quad$ Fermion $\quad$ &$\quad
\dfrac{1}{(2\pi)^{3/2}}\,\sqrt{\dfrac{m}{E_p}}\,u(p,\sigma)\quad$ &$\quad
\dfrac{1}{(2\pi)^{3/2}}\,\sqrt{\dfrac{m}{E_p}}\,\bar u(p,\sigma)\quad$ \\
&&\\
$\quad$ Antifermion $\quad$ &$\quad
\dfrac{1}{(2\pi)^{3/2}}\,\sqrt{\dfrac{m}{E_p}}\,\bar v(p,\sigma)\quad$ &$\quad
\dfrac{1}{(2\pi)^{3/2}}\,\sqrt{\dfrac{m}{E_p}}\,v(p,\sigma)\quad$ \\
&&\\
\hline
\end{tabular}
\end{center}
\end{table}

\noindent
7. For each change of the relative order of external fermions, a factor
$$(-1)\ .$$

For illustrative purposes, let's consider electron--positron scattering in
lowest non-trivial order of the perturbative, that is, loop, expansion (cf.
Fig.~\ref{fig:elposscat})\index{electron--positron scattering|(}:
$$
e^-(p_1,\sigma_1) + e^+(p_2,\sigma_2) \rightarrow
e^-(q_1,\tau_1) + e^+(q_2,\tau_2) \ .
$$
Applying the above Feynman rules, we may immediately write down the
corresponding scattering amplitude $T_{\rm f\/i}$ (in the so-called Feynman
gauge, defined by fixing the gauge parameter $\xi$ to the value $\xi = 1$):
\begin{eqnarray*}
T_{\rm f\/i} = &-& \frac{1}{(2\pi)^6}\,
\frac{m^2}{\sqrt{E_{p_1}\,E_{p_2}\,E_{q_1}\,E_{q_2}}}\,e^2\\[1ex]
&& \times\left[\frac{1}{(p_1 - q_1)^2}\,
\bar u(q_1,\tau_1)\,\gamma_\mu\,u(p_1,\sigma_1)\,
\bar v(p_2,\sigma_2)\,\gamma^\mu\,v(q_2,\tau_2)\right.\\[1ex]
&& \hspace{15pt} - \left.\frac{1}{(p_1 + p_2)^2}\,
\bar u(q_1,\tau_1)\,\gamma_\mu\,v(q_2,\tau_2)\,
\bar v(p_2,\sigma_2)\,\gamma^\mu\,u(p_1,\sigma_1)\right]\ .
\end{eqnarray*}
\begin{figure}[htb]
\normalsize
\begin{center}
\begin{tabular}{c}\\
\eexch\\[5ex]
(a)\\[6ex]
\eann\\[5ex]
(b)
\end{tabular}
\end{center}
\large
\caption{Electron--positron scattering in lowest order of the perturbation
expansion: (a) one-photon exchange, (b) pair annihilation.}\index{one-photon
exchange}\index{pair annihilation}\label{fig:elposscat}
\end{figure}
\index{electron--positron scattering|)}\index{Feynman rules|)}\index{gauge
theory|)}

\chapter{Quantum Chromodynamics}\index{quantum
chromodynamics|(}\index{QCD|(}\label{ch:qcd}

Quantum chromodynamics (QCD) is that quantum field theory which is generally
believed to describe the strong interactions. It is the special case of a
general gauge theory, characterized by the following features:

\begin{itemize}
\item The gauge group is SU(3)$_{\rm C}$ (where C stands for {\em colour}),
describing the (unbroken) symmetry acting on the colour degrees of freedom.
The order or dimension of the Lie group SU($N$) is $N^2 - 1$, which equals 8
in the case of SU(3). Hence, SU(3) has eight generators $T^a$, $a =
1,2,\dots,8$. Of course, the totally antisymmetric structure constants
$f_{abc}$, $a,b,c = 1,2, \dots ,8$, are those of SU(3); their values may be
found in Appendix~\ref{app:structureconstant}.
\item The particle content of QCD comprises the following vector-boson and
fermion fields:
\begin{itemize}
\item {\em Vector bosons\/}: There are eight, of course, massless, gluons
$G_\mu^a$ transforming according to the adjoint, i.e., eight-dimensional,
representation of SU(3).
\item {\em Fermions\/}: There are at least six quarks $q_f =
\mbox{u},\mbox{d},\mbox{s},\mbox{c},\mbox{b},\mbox{t},\dots ,$ each of them
transforming according to the fundamental, i.e., three-dimensional,
representation of SU(3). The generators of SU(3) in the fundamental
representation are given by
$$
T^a_{\rm fund} = \frac{\lambda_a}{2}\ ,
$$
where $\lambda_a$, $a = 1,2,\dots,8$, label the eight Gell-Mann matrices; an
explicit representation of the latter matrices may be found in
Appendix~\ref{app:gellmannmatrices}. The total number of quark flavours will
be denoted by $n_{\rm F}$: $f = 1,2,\dots,n_{\rm F}$.
\end{itemize}
\end{itemize}
Consequently, the Lagrangian defining QCD reads
\begin{eqnarray*}
{\cal L}^{\rm QCD} &=& -\frac{1}{4}\,F^a_{\mu\nu}\,F_a^{\mu\nu}
+ \sum_{f=1}^{n_{\rm F}}\bar q_f \left(i\,\Dsla - m_f\right)q_f\\
&+& \mbox{gauge-fixing terms}\\
&+& \mbox{ghost terms}\ ,
\end{eqnarray*}
with the gluon field strength
$$
F^a_{\mu\nu} = \partial_\mu G^a_\nu - \partial_\nu G^a_\mu
+ g_{\rm s}\,f_{abc}\,G^b_\mu\,G^c_\nu
$$
and the gauge-covariant derivative acting on the quark fields
$$
D_\mu = \partial_\mu - i\,g_{\rm s}\,G^a_\mu\,\frac{\lambda_a}{2} \ .
$$
The parameters of this theory are the strong coupling constant $g_{\rm s}$
and the (current) quark masses $m_f$.\index{quantum
chromodynamics|)}\index{QCD|)}

\chapter{SU(3)}\index{SU(3)|(}\label{app:gellmann}

\section{Structure constants}\index{structure
constants}\label{app:structureconstant}

The Lie group SU(3)\index{SU(3)} is defined by the following commutation
relations\index{commutator} for its eight
generators\index{generator}\index{commutator} $T^a$, $a = 1,\ldots ,8$:
$$
[T^a ,T^b] = i\,f_{abc}\,T^c\ ,
$$
where the nonvanishing elements among the structure constants $f_{abc}$ are
listed in Table~\ref{tab:fabc}.
\begin{table}[htb]
\caption{Nonvanishing structure constants $f_{abc}$ of SU(3)}\index{structure
constants}\label{tab:fabc}
$$
\begin{array}{cccr}\hline\hline\\[-1.5ex]
a&b&c&\multicolumn{1}{c}{f_{abc}}\\[1ex] \hline\\[-1.5ex]
1&2&3&1\\
1&4&7&1/2\\
1&5&6&\quad -1/2\\
2&4&6&1/2\\
2&5&7&1/2\\
3&4&5&1/2\\
3&6&7&\quad -1/2\\
4&5&8&\quad \sqrt{3}/2\\
6&7&8&\quad \sqrt{3}/2\\[1ex] \hline\hline
\end{array}
$$
\end{table}

In addition, the generators $T^a$ satisfy the anticommutation relations
$$
\{T^a,T^b\} = \frac{1}{3}\,\delta_{ab} + d_{abc}\,T^c\ ,
$$
where the nonvanishing elements among the coefficients $d_{abc}$ are listed
in Table~\ref{tab:dabc}.
\begin{table}[htb]
\caption{Nonvanishing coefficients $d_{abc}$ of
SU(3)}\label{tab:dabc}
$$
\begin{array}{cccl}\hline\hline\\[-1.5ex]
a&b&c&\multicolumn{1}{c}{d_{abc}}\\[1ex] \hline\\[-1.5ex]
1&1&8&\quad\msp 1/\sqrt{3}\\
1&4&6&\quad\msp 1/2\\
1&5&7&\quad\msp 1/2\\
2&2&8&\quad\msp 1/\sqrt{3}\\
2&4&7&\quad -1/2\\
2&5&6&\quad\msp 1/2\\
3&3&8&\quad\msp 1/\sqrt{3}\\
3&4&4&\quad\msp 1/2\\[1ex] \hline\hline
\end{array}
\quad \quad
\begin{array}{cccl}\hline\hline\\[-1.5ex]
a&b&c&\multicolumn{1}{c}{d_{abc}}\\[1ex] \hline\\[-1.5ex]
3&5&5&\quad\msp 1/2\\
3&6&6&\quad -1/2\\
3&7&7&\quad -1/2\\
4&4&8&\quad -1/(2\sqrt{3})\\
5&5&8&\quad -1/(2\sqrt{3})\\
6&6&8&\quad -1/(2\sqrt{3})\\
7&7&8&\quad -1/(2\sqrt{3})\\
8&8&8&\quad -1/\sqrt{3}\\[1ex] \hline\hline
\end{array}
$$
\end{table}
\noindent

The structure constants $f_{abc}$ are totally antisymmetric whereas the
coefficients $d_{abc}$ are totally symmetric under permutations of indices.

\section{Gell-Mann matrices}\index{Gell-Mann
matrices|(}\label{app:gellmannmatrices}

In the fundamental, i.e., three-dimensional, representation of SU(3), the
eight generators $T^a$, $a = 1,\ldots ,8$, are explicitly given by
$$
T^a_{\rm fund} = \dfrac{\lambda_a}{2}\ ,
$$
where $\lambda_a$ are the eight Gell-Mann matrices
$$
\begin{array}{c}
\lambda_1 = \left(\begin{array}{rrr}
0&1&0\\ 1&0&0\\ 0&0&0
\end{array}\right)\ ,\quad
\lambda_2 = \left(\begin{array}{rrr}
0&-i&0\\ i&0&0\\ 0&0&0
\end{array}\right)\ ,\quad
\lambda_3 = \left(\begin{array}{rrr}
1&0&0\\ 0&-1&0\\ 0&0&0
\end{array}\right)\ ,\\[4.5ex]
\lambda_4 = \left(\begin{array}{rrr}
0&0&1\\ 0&0&0\\ 1&0&0
\end{array}\right)\ ,\quad
\lambda_5 = \left(\begin{array}{rrr}
0&0&-i\\ 0&0&0\\ i&0&0
\end{array}\right)\ ,\\[4.5ex]
\lambda_6 = \left(\begin{array}{rrr}
0&0&0\\ 0&0&1\\ 0&1&0
\end{array}\right)\ ,\quad
\lambda_7 = \left(\begin{array}{rrr}
0&0&0\\ 0&0&-i\\ 0&i&0
\end{array}\right)\ ,\\[4.5ex]
\lambda_8 = \dfrac{1}{\sqrt{3}}\left(\begin{array}{rrr}
1&0&0\\ 0&1&0\\ 0&0&-2
\end{array}\right)\ .
\end{array}
$$

\section{Traces}\index{Gell-Mann matrices!---,
traces}\index{traces}\label{app:gmm-tr}

The traces of the simplest products of Gell-Mann matrices read
$$
\begin{array}{c}
\mbox{Tr}(\lambda_a )=0\ ,\\[1.5ex]
\mbox{Tr}(\lambda_a\,\lambda_b)=2\,\delta_{ab}\ ,\quad
\displaystyle\sum_a\mbox{Tr}(\lambda_a^2)=16\ ,\\[1.5ex]
\mbox{Tr}(\lambda_a\,[\lambda_b,\lambda_c])=4\,i\,f_{abc}\ ,\\[1.5ex]
\mbox{Tr}(\lambda_a\,\{\lambda_b,\lambda_c\})=4\,d_{abc}\ ,\\[1.5ex]
\mbox{Tr}(\lambda_a\,\lambda_b\,\lambda_c)=2\,i\,f_{abc}+2\,d_{abc}\ .
\end{array}
$$
Some further useful relations are:
$$
\begin{array}{c}
\displaystyle\sum_{a,b,c}(f_{abc})^2 =24\ ,\\[2.5ex]
\displaystyle\sum_{a,b,c}(d_{abc})^2 =\dfrac{40}{3}\ ,\\[2.5ex]
\displaystyle\sum_{j,k,a}\epsilon_{ijk}\,
\dfrac{\lambda^a_{\ell j}}{2}\,\dfrac{\lambda^a_{km}}{2}
= -\dfrac{2}{3}\,\epsilon_{i\ell m}\ .
\end{array}
$$
\index{Gell-Mann matrices|)}\index{SU(3)|)}

\pagestyle{myheadings}

\chapter{$\nabla_i\nabla_j\Phi(r)$}\label{app:nabla2}

\markboth{APPENDIX E. \ $\nabla_i\nabla_j\Phi(r)$}{}

We would like to express the second derivatives $\nabla_i\nabla_j\Phi(r)$
with respect to Cartesian coordinates ${\bf x}\equiv\{x_1,x_2,x_3\}$ of an
arbitrary function $\Phi(r)$ which depends merely on the radial coordinate
$r\equiv |{\bf x}|$ in terms of the ``spherically symmetric'' derivatives
coming into question, that is, the first and second derivatives of $\Phi$
with respect to $r$---which we indicate by prime(s)---as well as the
Laplacian $\Delta \equiv \mbox{\boldmath$\nabla$}\cdot\mbox{\boldmath
$\nabla$}$ of $\Phi$. In other words, we would like to rewrite these second
derivatives $\nabla_i\nabla_j\Phi(r)$ in terms of $\Phi'(r)$, $\Phi''(r)$,
and $\Delta\Phi(r)$. To this end, we start from the most general ansatz for
the expression we are looking for, viz., from
\begin{eqnarray*}
\nabla_i\nabla_j\Phi(r)
&=& \left(a\,\delta_{ij} + b\,\frac{x_i\,x_j}{r^2}\right)\frac{1}{r}\,\Phi'(r)
\\[1ex]
&+& \left(c\,\delta_{ij} + d\,\frac{x_i\,x_j}{r^2}\right)\Phi''(r)\\[1ex]
&+& \left(e\,\delta_{ij} + f\,\frac{x_i\,x_j}{r^2}\right)\Delta\Phi(r)\ ,
\end{eqnarray*}
where, for every term, the powers of the radial coordinate $r$ have been
chosen in such a way that the coefficients $a,b,\dots ,f$ are dimensionless.

It is a simple and straightforward task to determine the coefficients
$a,b,\dots ,f$:
\begin{itemize}
\item On the one hand, we contract the above ansatz by multiplying it by
$\delta_{ij}$ and by summing over $i$ and $j$. Using
$$
\delta_{ij}\,\delta_{ij} = \delta_{ij}\,\delta_{ji} = \mbox{Tr}(1_{3 \times
3}) = 3\ ,
$$
we obtain
$$
\Delta\Phi(r)
= (3\,a + b)\,\frac{1}{r}\,\Phi'(r)
+ (3\,c + d)\,\Phi''(r)
+ (3\,e + f)\,\Delta\Phi(r)\ .
$$
By comparing both sides of this equation, we may conclude that the
coefficients $a,b,\dots,f$ have to satisfy the relationships
\begin{eqnarray*}
3\,a + b &=& 0\ ,\\[1ex]
3\,c + d &=& 0\ ,\\[1ex]
3\,e + f &=& 1\ ,
\end{eqnarray*}
which, in turn, imply
\begin{eqnarray*}
a &=& -\frac{b}{3}\ ,\\[1ex]
c &=& -\frac{d}{3}\ ,\\[1ex]
e &=& \frac{1}{3} - \frac{f}{3}\ .
\end{eqnarray*}
Consequently, taking into account these relations and combining corresponding
terms, our ansatz simplifies to
\begin{eqnarray*}
\nabla_i\nabla_j\Phi(r)
&=& \left(\frac{x_i\,x_j}{r^2} - \frac{1}{3}\,\delta_{ij}\right)
\left[b\,\frac{1}{r}\,\Phi'(r) + d\,\Phi''(r) + f\,\Delta\Phi(r)\right]\\[1ex]
&+& \frac{1}{3}\,\delta_{ij}\,\Delta\Phi(r)\ .
\end{eqnarray*}
\item On the other hand, for the case $i \neq j$, we may easily calculate any
second derivative $\nabla_i\nabla_j\Phi(r)$ explicitly. The first derivative
$\nabla_i\Phi(r)$ of $\Phi(r)$ with respect to any of the Cartesian
coordinates $x_i$ reads
$$
\nabla_i\Phi(r) = \frac{x_i}{r}\,\Phi'(r)\ .
$$
Consequently, for $i \neq j$, the second derivatives $\nabla_i\nabla_j\Phi(r)$
of $\Phi(r)$ are given by
$$
\nabla_i\nabla_j\Phi(r) =
\frac{x_i\,x_j}{r^2}\left[\Phi''(r) - \frac{1}{r}\,\Phi'(r)\right]
\quad\mbox{for}\quad i \neq j\ ,
$$
whereas, by gaining advantage from the fact that $\delta_{ij} = 0$ for $i
\neq j$, the above, already simplified expression for
$\nabla_i\nabla_j\Phi(r)$ reduces to
$$
\nabla_i\nabla_j\Phi(r) =
\frac{x_i\,x_j}{r^2}\left[b\,\frac{1}{r}\,\Phi'(r) + d\,\Phi''(r)
+ f\,\Delta\Phi(r)\right]\quad\mbox{for}\quad i \neq j\ .
$$
The comparison of these two expressions allows us to fix the three, until now
indeterminate coefficients $b$, $d$, and $f$:
\begin{eqnarray*}
b &=& -1\ ,\\[1ex]
d &=& +1\ ,\\[1ex]
f &=& 0\ .
\end{eqnarray*}
\end{itemize}

In summary, upon collecting all our findings, the second derivatives
$\nabla_i\nabla_j\Phi(r)$ may be expressed as the following linear
combinations of the ``spherically symmetric'' derivatives $\Phi'(r)$,
$\Phi''(r)$, and $\Delta\Phi(r)$:
$$
\fbox{\ $\nabla_i\nabla_j\Phi(r) =
\left(\dfrac{x_i\,x_j}{r^2} - \dfrac{1}{3}\,\delta_{ij}\right)
\left[\Phi''(r) - \dfrac{1}{r}\,\Phi'(r)\right]
+ \dfrac{1}{3}\,\delta_{ij}\,\Delta\Phi(r)$\ }\ .
$$

\pagestyle{headings}

\chapter{Some Further Formulae for
Spectroscopy}\label{app:formspec}\index{Breit--Fermi Hamiltonian}

For unequal masses $m_1 \ne m_2$, the various spin-dependent relativistic
corrections\index{spin-dependent corrections}\index{relativistic
corrections}\index{corrections!---, spin-dependent}\index{corrections!---,
relativistic} to a nonrelativistic potential
$$
V_{\rm NR}(r) = V_{\rm V}(r) + V_{\rm S}(r)
$$
of vector--plus--scalar Lorentz structure read
\begin{itemize}
\item for the spin--orbit term\index{spin--orbit term}
\begin{eqnarray*}
H_{\rm LS}
&=&\dfrac{1}{4\,m_1^2\,m_2^2}\,\dfrac{1}{r}\left\{\dfrac{}{}\!\!\left[\left(
(m_1+m_2)^2+2\,m_1\,m_2\right){\bf L}\cdot{\bf S}_+\right.\right.\\[1ex]
&+&\left.\left(m_2^2-m_1^2\right){\bf L}\cdot{\bf S}_-\right]
\frac{d}{dr}V_{\rm V}(r)\\[1ex]
&-&\left.\left[\left(m_1^2 +m_2^2\right){\bf L}\cdot{\bf S}_+ +
\left(m_2^2 -m_1^2\right){\bf L}\cdot{\bf S}_-\right]
\frac{d}{dr}V_{\rm S}(r)\right\}\ ,
\end{eqnarray*}
with
$${\bf S}_+\equiv{\bf S}_1+{\bf S}_2$$ and
$${\bf S}_-\equiv{\bf S}_1-{\bf S}_2\ ;$$
\item for the spin--spin term\index{spin--spin term}
$$
H_{\rm SS}
= \frac{2}{3\,m_1\,m_2}\,{\bf S}_1\cdot{\bf S}_2\,\Delta V_{\rm V}(r)\ ;
$$
and
\item for the tensor term\index{tensor term}
$$
H_{\rm T}
= \frac{1}{12\,m_1\,m_2}\,S_{12}\left[\frac{1}{r}\,\frac{d}{dr}V_{\rm V}(r)
- \frac{d^2}{dr^2}V_{\rm V}(r)\right]\ .
$$
\end{itemize}
The signatures---{\em parity\/}\index{parity} $P$, {\em charge
conjugation\/}\index{charge conjugation} $C$, and {\em G parity\/}\index{G
parity}---for a quark--antiquark bound state with relative orbital angular
momentum $\ell$, spin $S$, and isospin $I$ are given by
\begin{eqnarray*}
P(q\bar q)&=&(-1)^{\ell +1}\ ,\\[1ex]
C(q\bar q)&=&(-1)^{\ell +S}\ ,\\[1ex]
G(q\bar q)&=&(-1)^{\ell +S+I}\ .
\end{eqnarray*}
For instance, for the pion we have $\ell=0$, $S=0$, $I=1$ and therefore
\begin{eqnarray*}
P(\pi)&=&-1\ ,\\[1ex]
C(\pi)&=&+1\ ,\\[1ex]
G(\pi)&=&-1\ .
\end{eqnarray*}

\chapter{Various Potential Models}\index{potential model|(}\label{ch:varpot}
\begin{table}[h]
\normalsize
$$
\begin{array}{lc}\hline\hline
&\\
\mbox{Eichten et al.} \; \cite{eichten75}
&V_{\rm NR} = -\dfrac{4}{3}\,\dfrac{\alpha_{\rm s}}{r}+a\,r\\
&\\ \hline
&\\
\mbox{Quigg--Rosner} \; \cite{quigg77}
&V_{\rm NR}=A\ln(r/r_0)\\
&\\ \hline
&\\
\mbox{Richardson} \; \cite{richardson79}
&V_{\rm NR} = -\dfrac{4}{3}\,\dfrac{48\,\pi^2}{33-2\,n_{\rm F}}\,
\dfrac{1}{(2\pi)^3}\int d^3q\,\dfrac{\exp(i\,{\bf q}\cdot{\bf x})}
{{\bf q}^2\ln (1+{\bf q}^2/\Lambda^2)}\\
&\\ \hline
&\\
\mbox{Ono--Sch\"oberl} \; \cite{ono79}
&V_{\rm NR}=-b\exp(-r/c)+\left\{\begin{array}{cc}
-\dfrac{4}{3}\,\dfrac{\alpha_{\rm s}}{r}+d &\mbox{for}\quad r\le R_1\\[2ex]
a\,r &\mbox{for}\quad r\ge R_1\end{array}\right. \\
&\\
&R_1 = \sqrt{\dfrac{4\,\alpha_{\rm s}}{3\,a}}\ ,\quad
\delta^{(3)}({\bf x}) \rightarrow \dfrac{1}{4\pi\,r_0^2}\,
\dfrac{\exp(-r/r_0)}{r}\\
&\\
&V_{\rm S} = a\,r\ ,\quad V_{\rm V} = V_{\rm NR} - V_{\rm S}\\
&\\ \hline
&\\
\mbox{Martin} \; \cite{martin80} &V_{\rm NR}=A+B\,r^{0.1}\\
&\\ \hline
&\\
\mbox{Buchm\"uller et al.} \; \cite{buchmuller80}
&V_{\rm NR}=-\dfrac{4}{3}\,\dfrac{1}{(2\pi)^3}\int d^3q
\exp(i\,{\bf q}\cdot{\bf x})\,
\dfrac{4\pi\,\alpha_{\rm s}({\bf q}^2)}{{\bf q}^2}\\
&\\ \hline
&\\
\mbox{Falkensteiner et al.} \; \cite{falk83}
&V_{\rm NR}=-\dfrac{4}{3}\,
\dfrac{\alpha_{\rm s}}{r}\,\mbox{erf}(\sqrt{\pi}\,A\,r)
+ a\,r\\
&\\ \hline\hline
\end{array}
$$
\large
\end{table}

\newpage

\begin{table}[h]
\normalsize
$$
\begin{array}{lc}\hline\hline
&\\
\mbox{Sch\"oberl} \; \cite{schoberl86}
&V_{\rm NR}=-b\exp(-r/c)+\left\{\begin{array}{cc}
-\dfrac{4}{3}\,\dfrac{\alpha_{\rm s}}{r}\,\mbox{erf}(A\,r)+d
&\mbox{for}\quad r\le R_1\\[2ex]
a\,r&\mbox{for}\quad r\ge R_1\end{array}\right.\\
&\\
&V_{\rm S} = a\,r\ ,\quad V_{\rm V} = V_{\rm NR} - V_{\rm S}\\
&\\ \hline
&\\
\mbox{Flamm et al.} \; \cite{flamm87}
&V_{\rm V}=-\dfrac{4}{3}\,\dfrac{\alpha_{\rm s}}{(r+r_0)^{1.107}}\,(1-c)
+a\,r^{0.91}\,(1-d)\\[3ex]
&V_{\rm S}=-\dfrac{4}{3}\,\dfrac{\alpha_{\rm s}}{(r+r_0)^{1.107}}\,c
+a\,r^{0.91}\,d\\[3ex]
&V_{\rm NR} =V_{\rm V} + V_{\rm S}\\
&\\ \hline\hline
\end{array}
$$
\large
\end{table}
\index{potential model|)}

\printindex

\end{document}